\documentclass[sigplan,10pt]{acmart}
\renewcommand\footnotetextcopyrightpermission[1]{} 
\acmConference[Arxiv'22]{Arxiv}{February 21, 2022}{Cornell University}
\acmYear{2022}
\acmISBN{} 
\acmDOI{} 
\startPage{1}

\setcopyright{none}

\usepackage{booktabs}   
\usepackage{subcaption} 
\usepackage{multirow}
\usepackage{graphicx}
\usepackage{stmaryrd}
\usepackage{semantic}
\usepackage{url}
\usepackage{color}
\usepackage{listings}
\usepackage[capitalize,noabbrev]{cleveref}
\usepackage[utf8]{inputenc}
\lstdefinelanguage{futhark}
{
  morekeywords={
    do,
    else,
    for,
    fun,
    if,
    in,
    include,
    let,
    loop,
    struct,
    then,
    type,
    val,
    while,
    with,
    module,
    where,
    sort,
    multired,
    reverse,
    zip3,
    unzip3,
    copy,
    gather,
    withacc,
    sum
  },
  sensitive=true, 
  morecomment=[l]{--}, 
  morecomment=[s]{\{-}{-\}}, 
  morestring=[b]", 
  literate={\\}{\fn}{1} {->}{$\rightarrow$}{1} {<-}{$\leftarrow$}{1} {|>}{$\pipe$}{1},
}

\lstdefinelanguage{corefuthark}
{
  morekeywords={
    do,
    else,
    for,
    fun,
    if,
    in,
    include,
    let,
    loop,
    struct,
    then,
    type,
    val,
    while,
    with,
    module,
    where,
  },
  sensitive=true, 
  literate={\\}{\fn}{1} {->}{$\rightarrow$}{1} {<-}{$\leftarrow$}{1},
  moredelim=**[is][\color{red}]{@}{@},
  morecomment=[l]{--}, 
  morecomment=[s]{\{-}{-\}}, 
  morestring=[b]" 
}

\usepackage{xcolor}
\definecolor{eclipseBlue}{RGB}{42,0.0,255}
\definecolor{eclipseGreen}{RGB}{63,127,95}
\definecolor{eclipsePurple}{RGB}{127,0,85}

\newcommand{\kw}[1]{\mbox{\texttt{\bfseries{#1}}}}

\newcommand{\Map}{\kw{map}}
\newcommand{\fn}{\ensuremath{\lambda}}
\newcommand{\pipe}{\ensuremath{\triangleright}}

\newcommand{\FnU}[2]{\fn#1~\rightarrow #2}
\newcommand{\Reduce}{\kw{reduce}}

\newcommand{\Scan}{\kw{scan}}

\newcommand{\tang}[1]{\dot{#1}}
\newcommand{\adj}[1]{\overline{#1}}

\usepackage{mathtools}
\usepackage{amsmath,amsbsy}
\usepackage{mathdots}
\usepackage[skip=5pt]{caption}

\begin{document}

\title{AD for an Array Language with Nested Parallelism} 


\author{Robert Schenck, Ola Rønning, Troels Henriksen and Cosmin E. Oancea}
\affiliation{
  \institution{DIKU, University of Copenhagen, Denmark}           
}
\email{r@bert.lol, ola@di.ku.dk, athas@sigkill.dk, cosmin.oancea@diku.dk}


\begin{abstract}
  We present a technique for applying (forward and) reverse-mode
  automatic differentiation (AD) on a non-recursive second-order
  functional array language that supports {\em nested parallelism}
  and is primarily aimed at {\em efficient GPU execution}.

  The key idea is to eliminate the need for a ``tape'' by
  relying on redundant execution to bring into each new scope all
  program variables that may be needed by the differentiated code.
  %
  Efficient execution is enabled by the observation that
  perfectly-nested scopes do not introduce re-execution, and such
  perfect nests are produced by known compiler transformations,
  e.g., flattening.
  Our technique differentiates loops and
  bulk-parallel operators---such as map, reduce, histogram, scan,
  scatter---by specific rewrite rules, and aggressively optimizes
  the resulting nested-parallel code.
  We report an experimental evaluation that compares with established
  AD solutions and demonstrates competitive performance on nine common
  benchmarks from recent applied AD literature.
%
\end{abstract}



\maketitle
\pagestyle{plain} 

\enlargethispage{\baselineskip}

\section{Introduction}


Automatic differentiation (AD) is a practical way for computing
derivatives of functions that are expressed as programs.  AD of
sequential code is implemented in tools such as
Tapenade~\cite{Araya-Polo2004DFA},
ADOL-C~\cite{griewank1996algorithm}, and
Stalingrad~\cite{lambda-backprop}.  Modern deep learning is built on
array programming frameworks such as
Tensorflow~\cite{abadi2016tensorflow} and
PyTorch~\cite{paszke2019pytorch}, which provide implicitly parallel
bulk operations that support AD.

A largely-unsolved challenge is supporting AD for high-level parallel
languages~\cite{futhark-pldi,10.1145/3473593,Next-700-Layers} %
that permit arbitrary nesting of sequential and parallel constructs.
Such solutions may in principle act as a catalyst for prototyping and
training of more advanced machine learning (ML) models.


ML involves minimising a \emph{cost function}, which
typically has far more inputs than outputs.  The \emph{reverse mode}
of AD is the most efficient in such cases~\cite{baydin-AD-survey}, but
is challenging to implement because intermediate program values are
required by the differentiated code. The program must first run a
\emph{forward sweep} (a.k.a., primal trace) that stores intermediate
program states on the \emph{tape}, which is then read during the
\emph{return sweep}, which essentially runs the program in reverse,
and actually computes the derivative.

A significant amount of work has studied how to elegantly model
reverse mode AD as a compiler transformation, and how to hide the
tape under powerful programming abstractions such as (dynamic)
closures~\cite{lambda-backprop} and delimited
continuations~\cite{shift-reset-backprop}.  These abstractions are
not suited for efficient parallel execution on manycore hardware
such as GPUs.


This work is to our knowledge the first to demonstrate an efficient
GPU implementation of reverse mode AD as a compiler transformation
on a data-parallel language that supports nested parallelism by
means of higher-order array combinators (SOAC), such as map,
reduce(-by-index), scan.

A key difference is that related approaches save by default
all variables on the tape and support checkpointing annotations
as an optimization (of memory footprint).
However, in a nested-parallel context, the tape may give raise
to complex, irregular data-structures that are passed across
deep nests and that are challenging to implement efficiently
in regards to optimizing spatial (coalescing) and temporal locality.


In contrast, our modeling of the tape takes inspiration from the
idea~\cite{baydin-AD-survey} that applying reverse AD to a straight
line of side-effect-free code does not require any tape per se,
because all intermediate values remain available.  We expand this
idea to drive the code-transformation across lexical scopes by
requiring that whenever the return sweep enters a new scope $s$, it
first redundantly re-executes the forward sweep of $s$ in order to
bring all the needed variables into scope.

\enlargethispage{\baselineskip}

Our technique preserves the work and span asymptotics
because the recomputation overhead is at worst proportional
to the depth of the deepest nest of scopes, which is constant
for a given nonrecursive program. Moreover, perfectly-nested
scopes (other than loops) are guaranteed to not introduce
re-execution, hence the overhead can be minimized by classic
compiler transformations such as flattening nested 
parallelism~\cite{futhark-ppopp} and polyhedral-like
optimizations~\cite{PolyPluto1}. Since our tape is essentially
formed by the in-scope variables, scalars are efficiently
accessed from registers rather than global memory (tape), and
the code resulted from AD fully benefits from the existent
compiler-optimization repertoire.

In what sequential loops are concerned, the loop-variant variables
require checkpointing because the return sweep needs to execute
the iterations in reverse order. In addition, we exploit to some
extent the fundamental time-space tradeoff studied by Siskind and
Pearlmutter~\cite{divide-and-conq}, in a simple and practical way,
by allowing the user to annotate how many times a loop should
be strip-mined.\footnote{
Strip-mining $k$ times a loop of count $2^k$ requires only
$2 k\times$ more memory than the original program at
the expense of $k\times$ re-execution at worst. 
Of course, since the strip-mining factor is a user-defined
constant it cannot achieve the logarithmic time-space overhead.
} 


Having designed the glue that binds scopes together, we turn
our attention to deriving high-level rewrite rules for
differentiating the parallel operators of the language.
We achieve this by starting from the main
rewrite rule of the reverse mode\footnote{
An original-program statement $x = f(a)$ produces the
differentiated code 
$ \overline{a} \ \text{+}= \ \frac{\partial \ f(a)}{\partial \ a} \cdot \overline{x}$.
where $\overline{x}$ denotes the adjoint of program variable $x$.   
}
and extend it by applying reasoning that combines imperative
(dependence analysis, loop distribution) and functional thinking
(rewrite rules, recurrences as scans).

In particular, the simplest parallel operator, namely map, is the most
difficult one to translate, because its purely functional semantics
allows free variables to be freely read inside it, but reverse AD
replaces a read with an accumulation, which cannot be in general
represented as a combination of classical data-parallel constructs.
In this context we report optimizations related to turning
accumulators to more-specialized constructs, such as reductions
and generalized histograms~\cite{histo-sc20}, that can yield
speedups close to one order of magnitude at the application level
(e.g., GMM and LSTM).


Our overall contribution is an end-to-end algorithm of AD with support
for nested parallel combinators, implemented inside (a clone of) the
compiler for the Futhark language and aimed at GPU execution.
Our specific contributions are:

\begin{itemize}
\item A redundant-execution technique for reverse AD that
  (i) eliminates the need for tape,
 (ii) is guaranteed to not introduce re-execution for
      perfectly-nested scopes other than loops, and
(iii) practically supports the time-space tradeoff by
      means of loop stripmining.
\item A set of rewrite rules for differentiating higher-order
  parallel combinators, including uses of free variables.
\item A collection of optimisations that rewrite common-cases
  of accumulators to other specific language constructs that
  benefit from specialized code generation.
\item An experimental evaluation that demonstrates
  (i) sequential performance competitive with
      Tapenade~\cite{Araya-Polo2004DFA} on
      ADBench~\cite{Srajer2018Abo},
 (ii) GPU performance competitive with Enzyme~\cite{10.1145/3458817.3476165}
      on two of their applications, i.e., RSBench, XSBench, and
(iii) significant GPU speedups in comparison with
      PyTorch~\cite{paszke2019pytorch} on GMM, LSTM, and for the 
      sparse formulation of K-Means.
\end{itemize}

\section{Preliminaries}
\label{sec:prelim}
This section provides the gist of how automatic differentiation of
programs may be derived from its mathematical foundation. Given a
function $P : \mathbb{R}^a \rightarrow \mathbb{R}^d$, the (total) derivative
of $P$ at a point $\mathbf{x} \in \mathbb{R}^a$ (where $P$ is
differentiable) can be written as a function of $\mathbf{y} \in
\mathbb{R}^a$ in terms of its Jacobian as
$\frac{\partial{P(\mathbf{x})}}{\partial{\mathbf{x}}}(\mathbf{y}) \ =
\ J_P(\mathbf{x}) \cdot \mathbf{y}$, where $\cdot$ is matrix-vector
multiplication.

The Jacobian is a representation of a function derivative at a
specific point. The {\em chain rule} provides a straightforward way of
deriving the derivative across function composition. For example, if
$P(\mathbf{x}) = h(g(f(\mathbf{x})))$, and the types of $f$, $g$, $h$
are $f : \mathbb{R}^a \rightarrow \mathbb{R}^b, g : \mathbb{R}^b
\rightarrow \mathbb{R}^c, \text{ and } h : \mathbb{R}^c \rightarrow
\mathbb{R}^d$, then:
\vspace{-5pt}
\begin{equation}\label{eq:chain-rule}
J_P(\mathbf{x}) = J_{h} \left(g(f(\mathbf{x}))\right)\cdot
                  J_{g}\left(f(\mathbf{x})\right)\cdot
                  J_{f}(\mathbf{x}). \\[-5pt]
\end{equation}
Since matrix multiplication is associative, there are multiple ways to
evaluate $J_P(\mathbf{x})$ when it's viewed as a program.  For example,
from right-to-left:\smallskip\\
$\mbox{ }\hspace{10ex}
\overrightarrow{J_P}(\mathbf{x}) = J_{h}
\left(g(f((\mathbf{x}))\right)
\underbrace{\left[J_{g}\left(f(\mathbf{x})\right)
    J_{f}(\mathbf{x})\right]}_{\bigstar}, $\vspace{-2ex}\\ or
left-to-right:\vspace{-2ex}\\ $ \mbox{ }\hspace{10ex}
\overleftarrow{J_P}(\mathbf{x}) = \overbrace{\left[J_{h}
    \left(g(f(\mathbf{x}))\right)
    J_{g}\left(f(\mathbf{x})\right)\right]}^{\blacksquare}
J_{f}(\mathbf{x}).  $\vspace{2pt}

Notice that if $h$ is a scalar function (i.e., $d = 1$), then
$J_{h}\left(g(f(\mathbf{x})\right)$ is a gradient vector of dimension
$\mathbb{R}^{d}$. In the right-to-left formulation, $\bigstar$ is a
large matrix of dimension $\mathbb{R}^{c \times a}$, whereas in the
left-to-right formula, $\blacksquare$ is a vector of dimension
$\mathbb{R}^{b}$.  When $d \ll a$, it is much more expensive to
compute/manifest the larger intermediate Jacobian matrices in the
right-to-left evaluation order. When $a \ll d$, it's more efficient to
choose the right-to-left evaluation order. There's a catch: the
Jacobians depend on intermediate results, e.g., $J_g(f(\mathbf{x}))$
depends on $f(\mathbf{x})$. In the right-to-left evaluation order, the
computation of these intermediate results can be interwoven with the
computation of the Jacobians, as the Jacobians are computed in
program-order.  In the left-to-right evaluation there's no such luck:
$P(\mathbf{x})$ must first be executed, with all intermediate results
saved, before the $J_P(\mathbf{x})$ may be computed. For this reason,
when $a \approx d$ the right-to-left evaluation order is preferred.
\subsubsection{Forward mode AD}
\begin{scriptsize}
\begin{figure}
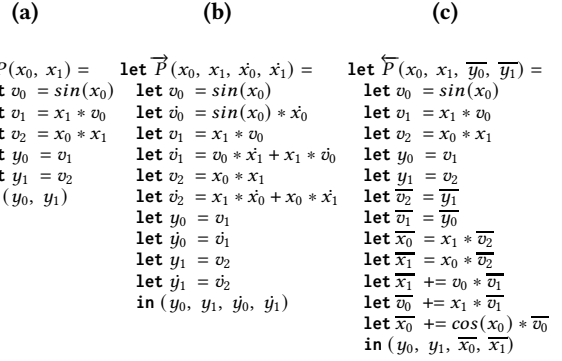

  \begin{subfigure}[t]{0.24\columnwidth}
  \caption{}
  \[
  \begin{array}{l}
    \kw{let} \ P (x_0, \ x_1) = \phantom{\overrightarrow{P}}\\
    \hspace{2ex} \kw{let} \ v_0 \ = sin(x_0)  \\
    \hspace{2ex} \kw{let} \ v_1 \ = x_1 * v_0 \\
    \hspace{2ex} \kw{let} \ v_2 \ = x_0 * x_1 \\
    \hspace{2ex} \kw{let} \ y_0 \ = v_1 \\
    \hspace{2ex} \kw{let} \ y_1 \ = v_2 \\
    \hspace{2ex} \kw{in} \ (y_0, \ y_1) \\
    \end{array}
  \]
  \label{fig:simple-ex}
\end{subfigure}
\begin{subfigure}[t]{0.35\columnwidth}
  \caption{}
  \[
  \begin{array}{l}
    \kw{let} \ \overrightarrow{P} (x_0, \ x_1, \ \tang{x_0}, \ \tang{x_1}) = \\
    \hspace{2ex} \kw{let} \ v_0 \ = sin(x_0)  \\
    \hspace{2ex} \kw{let} \ \tang{v_0} \ = sin(x_0) * \tang{x_0}  \\
    \hspace{2ex} \kw{let} \ v_1 \ = x_1 * v_0 \\
    \hspace{2ex} \kw{let} \ \tang{v_1} \ = v_0 * \tang{x_1} + x_1 * \tang{v_0} \\
    \hspace{2ex} \kw{let} \ v_2 \ = x_0 * x_1 \\
    \hspace{2ex} \kw{let} \ \tang{v_2} \ = x_1 * \tang{x_0} + x_0 * \tang{x_1} \\
    \hspace{2ex} \kw{let} \ y_0 \ = v_1 \\
    \hspace{2ex} \kw{let} \ \tang{y_0} \ = \tang{v_1} \\
    \hspace{2ex} \kw{let} \ y_1 \ = v_2 \\
    \hspace{2ex} \kw{let} \ \tang{y_1} \ = \tang{v_2} \\
    \hspace{2ex} \kw{in} \ (y_0, \ y_1, \ \tang{y_0}, \ \tang{y_1}) \\
    \end{array}
  \]
  \label{fig:simple-tan-ex}
\end{subfigure}
  \begin{subfigure}[t]{0.35\columnwidth}
  \caption{}
  \[
  \begin{array}{l}
    \kw{let} \ \overleftarrow{P} (x_0, \ x_1, \ \adj{y_0}, \ \adj{y_1})= \\
    \hspace{2ex} \kw{let} \ v_0 \ = sin(x_0)  \\
    \hspace{2ex} \kw{let} \ v_1 \ = x_1 * v_0 \\
    \hspace{2ex} \kw{let} \ v_2 \ = x_0 * x_1 \\
    \hspace{2ex} \kw{let} \ y_0 \ = v_1 \\
    \hspace{2ex} \kw{let} \ y_1 \ = v_2 \\
    \hspace{2ex} \kw{let} \ \adj{v_2} \ = \adj{y_1} \\
    \hspace{2ex} \kw{let} \ \adj{v_1} \ = \adj{y_0} \\
    \hspace{2ex} \kw{let} \ \adj{x_0} \ = x_1 * \adj{v_2} \\
    \hspace{2ex} \kw{let} \ \adj{x_1} \ = x_0 * \adj{v_2} \\
    \hspace{2ex} \kw{let} \ \adj{x_1} \ \mathrel{+}= v_0 * \adj{v_1}\\
    \hspace{2ex} \kw{let} \ \adj{v_0} \ \mathrel{+}= x_1 * \adj{v_1}\\
    \hspace{2ex} \kw{let} \ \adj{x_0} \ \mathrel{+}= cos(x_0) * \adj{v_0} \\
    \hspace{2ex} \kw{in} \ (y_0, \ y_1, \  \adj{x_0}, \ \adj{x_1}) \\
    \end{array}
  \]
  \label{fig:simple-adj-ex}
\end{subfigure}\vspace{-3ex}
  \caption{{\bf (a)} A program $P$ which computes the function ${f(x_0,x_1) = (x_1 \cdot sin(x_0),x_0 \cdot x_1)}$,
           {\bf (b)} the forward mode AD transformation of the program, and
           {\bf (c)} the reverse mode AD transformation of the program.\vspace{-13pt}}
\end{figure}
\end{scriptsize}
Consider the example program $P$ in \cref{fig:simple-ex}.  The
\emph{forward mode} AD transformation of $P$, $\overrightarrow{P}$, is
shown in \cref{fig:simple-tan-ex}; the transformation involves
the core rewrite rule:\footnote{This rule generalizes for $n$-array
functions in the obvious way: for any statement $\kw{let} \ v =
f(a_0,\dots,a_{n-1})$, we have $\tang{v} =
\sum_{i=0}^{n-1}\frac{\partial f}{\partial a_i} \tang{a_i}$.}\vspace{-1ex}
\begin{equation} \label{eq:rewrite-tan}
\kw{let} \ v = f(a,b) \ \Longrightarrow 
\begin{aligned}
&\kw{let} \ v = f(a,b) \\[-4pt]
&\kw{let} \ \tang{v} = \frac{\partial f(a,b)}{\partial a} \tang{a} + \frac{\partial f(a,b)}{\partial b} \tang{b}
\end{aligned}
\end{equation}
where we call $\tang{v}$ the \emph{tangent} of $v$, which is the
derivative of $v$ with respect to a chosen direction $(\tang{x_0},
\ \tang{x_1})$ (now appearing as an additional argument to
$\overrightarrow{P}$).  $\overrightarrow{P}$ also returns the tangents
of its outputs, $\tang{y_0}$ and $\tang{y_1}$, which make-up the
derivative of $P$.
We expose forward mode AD to the user via a function
\vspace{-5pt}
\[
jvp : (P: \mathbb{R}^n \rightarrow \mathbb{R}^m) 
      \rightarrow (\overrightarrow{P} : \mathbb{R}^n \rightarrow \mathbb{R}^{n} \rightarrow \mathbb{R}^m) \\[-5pt]
\] which, for any differentiable program $P$ computes
$\overrightarrow{P}$, with the property that $(\overrightarrow{P}
(\mathbf{x})) (\tang{\mathbf{x}}) = \overrightarrow{J}_P(\mathbf{x})
\tang{\mathbf{x}}$.
\footnote{$jvp$ stands for \emph{Jacobian-vector product}, named as
such because we apply the direction vector on the right of the
Jacobian.}  In simpler terms, $\overrightarrow{P}$ takes a point and a
direction and returns the derivative of $P$ at the point and in the
given direction by multiplying the direction on the right of the
right-to-left association of the Jacobian for
$P$.  In order to recover the full Jacobian
$\overrightarrow{J}_{P}(\mathbf{x})$, $\overrightarrow{P}(\mathbf{x})$
is mapped over the standard basis of $\mathbb{R}^n$.

\subsubsection{Reverse mode AD}
\label{subsubsec:rev-ad}
The \emph{reverse mode} AD transformation of $P$, $\overleftarrow{P}$,
is shown in \cref{fig:simple-adj-ex}. As with the forward mode
transformation, it involves the core rewrite rule:\footnote{
In general, the adjoint of a variable $a$ appearing on the RHS
of statements $\kw{let} \ v_0 = f_0(.., a,..), \dots, \  \kw{let}\ v_{n-1} = f_{n-1}(.., a,..)$
is $\adj{a} = \sum_{i=0}^{n-1} \frac{\partial f}{\partial a} \adj{v_i}$.}
\begin{equation} \label{eq:rewrite-adj}
\kw{let} \ v = f(a,b) \ \Longrightarrow
\begin{aligned}
&\kw{let} \ v = f(a,b) \\[-8pt]
&\hspace{7pt}\mbox{\scriptsize$\vdots$}\\[-12pt]
&\kw{let} \ \adj{a} \mathrel{+}= \frac{\partial f(a,b)}{\partial a} \adj{v} \\[-3pt]
&\kw{let} \ \adj{b} \mathrel{+}= \frac{\partial f(a,b)}{\partial b} \adj{v} 
\end{aligned}
\end{equation}
where $\adj{v}$ is the \emph{adjoint} of $v$, which is the derivative
of $P$ with respect to $v$. The adjoint of a variable $v$ may be
updated multiple times--any use of $v$ may contribute to the
derivative (see, e.g., $x_1$ in
\cref{fig:simple-adj-ex}). Reverse mode AD corresponds to computing the
left-to-right Jacobian $\overleftarrow{J_P}{(\mathbf{x})}$. All
intermediate results of $P$ which appear in adjoint expressions must
first be saved, which corresponds to computing $P$ before the adjoint
of each program variable. The $\smash{\vdots}$ indicates the presence
of these statements, along with any preceding adjoint
expressions. The adjoints in \cref{fig:simple-adj-ex} are
computed in the reverse program order, corresponding to a
left-to-right association of the Jacobian of $P$. Any adjoint
$\adj{v}$ is necessarily finalized before it is used: all uses of $v$
must occur after assignment; any contributions to $\adj{v}$ must
necessarily appear before any uses of $\adj{v}$ in reverse program
order.
$\overleftarrow{P}$ takes as additional arguments the initial
adjoints of the output $\adj{y_0},\  \adj{y_1}$ and additionally returns
the adjoints of the input, $\adj{x_0},\ \adj{x_1}$.
Reverse mode AD is made available to the user via a function:
\vspace{-5pt}
\[
vjp : (P: \mathbb{R}^n \rightarrow \mathbb{R}^m) 
      \rightarrow (\overleftarrow{P} : \mathbb{R}^n \rightarrow \mathbb{R}^{m} \rightarrow \mathbb{R}^n) \\[-5pt]
\]
which computes $\overleftarrow{P}$ for any differentiable program
$P$, with the property that $(\overleftarrow{P} (\mathbf{x}))
(\adj{\mathbf{y}}) = \adj{\mathbf{y}}\overleftarrow{J}_P(\mathbf{x})$.
\footnote{$vjp$ stands for \emph{vector-Jacobian product}.} As
$\overleftarrow{P}$ has the same computational properties as
$\overleftarrow{J}_P(\mathbf{x})$, it is the preferred choice when $m
\ll n$. When $P$ computes a scalar function (i.e., $m = 1$), $vjp$
returns the complete gradient of $P$ in a single pass; $jvp$ requires
$n$ passes, one for each basis vector of $\mathbb{R}^n$.  Programs
with very high-dimensional inputs and low-dimensional outputs appear
frequently in the fields of optimization and machine learning, making
$vjp$ the computationally superior choice. The $vjp$ transformation is
also more complex due to computing adjoints in reverse order, storing
of intermediate program variables, and accumulation of derivatives
into adjoint variables, which turn any read of a variable in the
original program into a write in the transformed program.  For these
reasons, we concentrate our discussion on reverse mode AD.

Higher-order derivatives are supported by nesting of
$vjp$ or $jvp$.

\subsection{Language}
\label{sec:language}

We perform our transformation on a data-parallel language which
features arbitrary nesting of \emph{second-order array combinators}
(SOACs), e.g., $\Map$, $\Reduce$, and $\Scan$.  SOACs are variadic
in their number of arguments and returns, i.e., \kw{zip}s
and \kw{unzip}s are implicit. For example,
\vspace{-5pt}
\[
\kw{unzip3}~(\kw{map}~(\FnU{(a,b,c)}{(c, b, a)})~(\kw{zip3}~as~bs~cs))\\[-5pt]
\]
may be equivalently written as simply
\vspace{-5pt}
\[
\kw{map}~(\FnU{a~b~c}{(c, b, a)})~as~bs~cs\\[-5pt]
\]

The source language supports higher-order functions, polymorphism,
modules, and similar high-level features, which are compiled away
using a variety of
techniques~\cite{Elsman:2018:SIH:3243631.3236792,tfp18hovgaard} before
we perform AD.  The only remaining second-order functions are the
SOACs. Lambdas can only appear syntactically in SOACs and VJP/JVP,
and are not values. As such we do not suffer from 
``perturbation confusion''~\cite{perturb-confusion}.
Further, a significant battery of standard optimisations (CSE,
constant folding, aggressive inlining) is also applied prior to AD.

The language is written in A-normal form~\cite{Sabry:1992:RPC:141478.141563}
(ANF): all subexpressions are
variables names or constants with the exception of the body expression of
$\kw{loops}$, $\kw{if}$-expressions and $\kw{let}$-expressions.
$\kw{let}$-expressions consist of a series of bindings---that we
also call statements---followed by a sequence of one
or more returns:
\vspace{-5pt}
\[\kw{let}~a~=~5~*~x~~\kw{let}~b~=~a~*~a~~\kw{in}~b\]
The language is purely functional: re-definitions of the same variable
(in a given scope) should be understood as a notational
convenience for variable shadowing. It
supports a functional flavor of in-place updates based on
uniqueness types~\cite{futhark-pldi}.
The binding $\kw{let} \ xs[i] = \ x$ is syntactic sugar for $\kw{let} \ xs'
\ = \ xs \ \kw{with} \ [i] \leftarrow x$, which has the semantics
that $xs'$ is a copy of $xs$ in which the element at index $i$
is updated to $x$, but also provides the operational guarantee
that the update will be realized in place.

The language also features sequential and pure loops, which have the
semantics of a tail-recursive function:
\vspace{-5pt}
\[
\kw{let}~y~=~\kw{loop}~(x)=(x_0)~\kw{for}~i=0\ldots~n-1~\kw{do}~e\\[-5pt]
\]
The loop is initialized by binding $x_0$ to $x$. Each iteration of the
loop executes $e$ , and binds the result of the expression to $x$,
which is used on the subsequent iteration of the loop. The loop
terminates after $n$ iterations and the final result of $e$ is bound
to $y$. We introduce the remainder of the language on a by-need basis.

\section{Forward mode}
We implement forward mode AD by
straight-forward application of the forward mode rewrite rule
(\ref{eq:rewrite-tan}). That is, tangent variables statements are
interleaved with primal variable statements (tangent variables are
associated with primal variables via a simple mapping). In parameters,
body results, and SOAC arguments we bundle tangent variables
together with their primal counterpart.

As an alternative to a program transformation, forward mode AD can be
implemented by coupling primal and tangent variables together in a
tuple, e.g., $(v, \tang{v}) = (f(a), \frac{d f}{d a} \tang{a})$.  Via
overloading, the operators in the program may operate directly on such
tuples, giving rise to the operator-overloading/ dual number
formulation of forward-mode AD described in
\cite{baydin-AD-survey}. Our approach of primal and tangent statements
and bundling of primal and tangent variables is equivalent to
this formulation of AD, implemented as a compiler pass instead of via
operator overloading.

\section{Reverse-AD by Redundant Execution}
\label{sec:tape-loops}

We first demonstrate the gist of our technique on
a perfect-nest example in \cref{subsec:gist-perfect-nests},
then present the code transformation across scopes in a
semi-formal manner in~\cref{transf-scopes}, and conclude
with an example demonstrating our code generation for
loops and the time-space trade-off in~\cref{subsec:main-tradeoff}.

\subsection{The Gist of Our Technique}
\label{subsec:gist-perfect-nests}

\begin{figure} 
\begin{lstlisting}[mathescape,basicstyle=\ttfamily\scriptsize] 
let xss= map(\c as -> -- fwd re-execution scope 0
              if c then ... else map(\a -> a*a) as) cs ass
let $\overline{ass}$ = -- return sweep
  map(\c as $\overline{xs}$ ->
       let xs = if c then ... -- fwd re-execution scope 1
                else map(\a -> a*a) as
       in  if c then ...      -- reverse sweep scope 1
                else -- fwd re-execution scope 2
                  let xs' = map (\a -> a*a) as
                  let $\overline{as}$ = map (a $\overline{a}$ $\overline{x}$ ->
                               -- fwd re-execution scope 3
                                 let x = a * a
                                 in  2 * a * $\overline{x}$
                            ) as $\overline{xs}$
                  in  $\overline{as}$
     ) cs ass $\overline{xss}$
in  $\overline{ass}$
\end{lstlisting}\vspace{-2ex}
\caption{ Figure shows the reverse-AD code of the program:\newline
          \lstinline{map(\\c as -> if c then ... else map (\a->a*a) as) cs ass}\newline
          Variables {\tt xss}, {\tt xs'}, {\tt xs'} and {\tt x} correspond
          to the result of forward-sweep re-execution of scopes $0$, $1$, $2$ and $3$.
          Since the program is perfectly nested, these variables
          are guaranteed to be dead-code, i.e., no re-execution
          overhead is introduced.
        }
\label{fig:perfect-nest}
\end{figure}

Figure~\ref{fig:perfect-nest} shows the code generated by applying
reverse-mode AD to the function below, and to the result adjoint 
$\overline{xss}$:\footnote{
Map independently applies a function argument to each of the
elements of an array (argument), 
i.e., $\kw{map}~f~[x_1,\ldots,x_n]~=~[f x_1,\ldots, f x_n]$.
}\vspace{-0.5ex}

\begin{lstlisting}[mathescape,basicstyle=\ttfamily\scriptsize,float=h]
\ ass -> let xss = map(\c as -> let xs = if c then ...
                                   else map(\a -> a*a) as
                                in xs
                      ) cs ass
         in  xss
\end{lstlisting}\vspace{-0.5ex}

The program has four scopes: the outermost one (scope 0),
the one of the outer-map lambda (scope 1), the one of the
{\tt then/else} branches (scope 2), and the one of the
inner-most lambda (scope 3).  While generating the return
sweep, our technique re-executes the forward sweep of the
corresponding scope ($1$, $2$ and $3$) in order to make
available the original program variables that might be
used in the differentiated code. Essentially, the forward
sweep is executed at worst four times, i.e., the depth of
the deepest program nest. 

However, it is easy to observe that perfect nests (other
than loops) are guaranteed to not introduce re-computation,
because, by definition, their bodies consists of
one (composed) statement, hence the scope result cannot
possibly be used by the return sweep.
In \cref{fig:perfect-nest} the bindings of {\tt xss},
{\tt xs}, {\tt xs'} and {\tt x} are all dead code, hence
the differentiated program after optimization
suffers no re-computation overheads:\vspace{-0.5ex}
\begin{lstlisting}[mathescape,basicstyle=\ttfamily\scriptsize]
let $\overline{ass}$ = map(\c as $\overline{xs}$ -> if c then ... else
                    map(\a $\overline{x}$ -> 2*a*$\overline{x}$) as $\overline{xs}$ ) cs ass $\overline{xss}$
\end{lstlisting}\vspace{-0.5ex}

It follows that overheads can be optimized by
known transformations~\cite{PolyPluto1, PolyPluto2} rooted
in loop distribution and interchange, also supported by 
Futhark~\cite{futhark-ppopp}. With their help, we commonly
expect the forward sweep to be executed twice: 
\begin{itemize}
\item[1] once for the outermost scope, as expected, because
  programs typically consist of multiple nests and also
  the user may require the result of the original program,
\item[2] once for the innermost scope that typically performs
  scalar computation, which are cheap to recompute. In comparison,
  vectorization or the use of the tape would require such scalars
  to be retrieved from global memory, which is about
  two order of magnitude slower.
\end{itemize}

%

\subsection{Transformation Rules Across Scopes}
\label{transf-scopes}

\newcommand{\FV}{\mathcal{FV}}

\begin{figure}
\begin{scriptsize}
  \textbf{Rule $vjp_{body}$ refers to a body $body = stms \ \kw{in} \ res$:}\vspace{-1ex}
    \[
      \begin{array}{l}
        vjp_{body}(\overline{res}, \ stms \ \kw{in} \ res) \Rightarrow 
            \overleftrightarrow{stms} \ \kw{in} \ (res, \ \overline{fvs_{body}}) \\
        \mbox{\tt~} \kw{where} \ \
        \overleftrightarrow{stms} \leftarrow vjp_{stms}(stms) \ \ \kw{and} \ \
        \overline{fvs_{body}} \leftarrow \overline{\FV(body)}
      \end{array}
    \]\medskip

  \textbf{Rule $vjp_{\lambda}$ refers to a lambda function $\lambda x_1\ldots x_n \rightarrow body$:}\vspace{-1ex}
    \[
      \begin{array}{l}
        vjp_{\lambda}(\overline{res}, \ \lambda x_1\ldots x_n \rightarrow stms \ \kw{in} \ r) \Rightarrow
            \lambda x_1\ldots x_n \rightarrow \overleftrightarrow{body} \\
        \mbox{\tt~} \kw{where} \ \
                     \overleftrightarrow{stms} \ \kw{in} \ (\_, \overline{fvs_{body}}) \leftarrow vjp_{body}(\overline{res}, \ body) \\
        \hspace{8ex} \overleftrightarrow{body} \leftarrow \overleftrightarrow{stms} \ \kw{in} \ \overline{fvs_{body}}
      \end{array}
    \]\medskip

  \textbf{Rule $vjp_{stms}$ simply folds over each statement:}\vspace{-1ex}
    \[
      \begin{array}{l}
        vjp_{stms}(stm, stms) \Rightarrow \overrightarrow{stms}, \ vjp_{stms}(stms), \ \overleftarrow{stms} \\
        \mbox{\tt~} \kw{where} \ \ \
            (\overrightarrow{stms}, \overleftarrow{stms}) \leftarrow vjp_{stm}(stm)
      \end{array}
    \]\medskip


  \textbf{Rule $vjp_{stm}$ for scalar multiplication (simple) statement:}\vspace{-1ex}
    \[
      \begin{array}{l}
        vjp_{stm}(\kw{let} \ x \ = \ a \ * \ b) \Rightarrow (\overrightarrow{stms}, \ \overleftarrow{stms}) \\
        \mbox{\tt~} \ \ \kw{where} \ \ \
        \overrightarrow{stms} \leftarrow \kw{let} \ x \ = \ a \ * \ b \\
        \hspace{10ex} \overleftarrow{stms} \leftarrow \kw{let} \ \overline{a} \ = \ \overline{a} \ + \ b * \overline{x}, \ \ \kw{let} \ \overline{b} \ = \ \overline{b} \ + \ a * \overline{x}
      \end{array}
    \]\smallskip

 \textbf{Rule $vjp_{stm}$ for loop statement: checkpoints and re-executes body}\\
    (Applies only to loops whose variants
        do not change their shape within the loop)\vspace{-1ex}
    \[
      \begin{array}{l}
        vjp_{stm}(\kw{let} \ y \ = \ \kw{loop} \ (x)=(x_0) \ \kw{for} \ i=0\ldots n-1 \ \kw{do} \ body) 
            \Rightarrow (\overrightarrow{loop}, \ \overleftarrow{loop}) \\
        \hspace{3ex} \kw{where} \\
        \hspace{5ex} stms \ \kw{in} \ r \ \leftarrow \ body \\
        \hspace{5ex} \overrightarrow{loop} \leftarrow \kw{let} \ xs_0 = \ \kw{scratch}(n, \kw{size}(\kw{typeOf}(x))),\\
        \hspace{14ex} \kw{let} \ (y, xs) \ = \ \kw{loop} \ (x, xs) = (x_0, xs_0)\\
        \hspace{16ex} \kw{for} \ i = 0 \ldots n-1 \ \kw{do} \ \ 
                        \kw{let} \ xs[i] \ = \ x, \ stms \ \kw{in} \ (r, xs)\  \\
        \hspace{5ex} fvs_{bdy} \ \leftarrow \ \FV(body)\\
        \hspace{5ex} \overleftrightarrow{stms} \ \kw{in} \ (\_, \
                \overline{fvs_{bdy}'}) \ \leftarrow \ vjp_{body}(\overline{x}, \ body)\\
        \hspace{5ex} (\_, \ \overleftarrow{stm_{x_0}}) \leftarrow vjp_{stm}(\kw{let} \ x' \ = \ x_0)\\
        \hspace{5ex} \overleftarrow{loop} \leftarrow \kw{let} \ \overline{fvs_{bdy_0}} \ = \mbox{\tt get or initialize adjoints},\\
        \hspace{14ex} \kw{let} \ (\overline{x'}, \ \overline{fvs_{bdy}}) \ = \\
        \hspace{16ex} \kw{loop} \ (\overline{x}, \overline{fvs_{bdy}}) = (\overline{y}, \ \overline{fvs_{bdy_0}})\\
        \hspace{18ex} \kw{for} \ i \ = \ n-1 ... 0 \ \kw{do}\\
        \hspace{20ex}   \kw{let} \ x \ = \ xs[i], \ \overleftrightarrow{stms}
                        \ \kw{in} \ (\overline{r}, \overline{fvs_{bdy}'}),\\
        \hspace{14ex} \overleftarrow{stm_{x_0}}
      \end{array}
    \]
\end{scriptsize}\vspace{-2ex}
  \caption{Reverse AD transformation rules for body of statements, scalar multiplication and do loops.}
  \label{fig:rules-loop}
\end{figure}
%
%

Figure~\ref{fig:rules-loop} sketches the reverse-mode AD code
transformation at the lexical levels of a body of statements
($vjp_{body}$), a sequence of statements ($vjp_{stms}$) and
individual statements ($vjp_{stm}$). We recall that {\em body}
refers to a sequence of $\kw{let}$ bindings (statements)
followed by an $\kw{in}$ result.
Our presentation sacrifices some formalism for readability, for
example, by omitting the environment, which essentially carries out a
mapping from the variables of the source program to their respective
adjoints. Instead, we assume that the adjoint $\overline{x}$
of a variable $x$ is always available.

The $vjp_{body}$ rule refers to transforming a body of statements,
which defines a new scope.
The rule starts by binding the body result to its adjoint $res \mapsto
\overline{res}$ (not shown). This is safe because the transformation
works backwards, hence $\overline{res}$ is already available from the
outer scope (ultimately passed as parameter to $vjp_{function}$ by the user).
The statements of the transformed body
$\overleftrightarrow{stms}$ are those generated by $vjp_{stms}$ and
the result consists of the original result $res$, extended with the
adjoints $\overline{fvs_{body}}$, consisting of all free variables
that are used inside the body---$\FV(body)$ finds the free variables
used in the body and $\overline{\FV(body)}$ refers to
their adjoints.

Rule $vjp_{\lambda}$ refers to an unnamed (lambda) function, and
its reverse-AD code is essentially obtained by calling $vjp_{body}$
on lambda's body. Note that $x_{1,\ldots,n}$ are free variables
in $body$, and as such their adjoints are among $\overline{fvs_{body}}$.

The $vjp_{stms}$ rule highlights the redundant-execution mechanism
that removes the need to implement the tape as a separate abstraction:
each statement $stm$ is processed individually (by $vjp_{stm}$),
producing a sequence of statements on the forward sweep, denoted
$\overrightarrow{stms}$, that brings into scope whatever information
is necessary to execute the return sweep for that statement, denoted
by $\overleftarrow{stms}$.\footnote{ The statements of the forward and
return sweeps are arranged symmetrically; all statements of
the forward sweep come before any of the return sweep, and the latter
is organized in the reverse order of the original statements:
$vjp_{stms}(stm, stms) \Rightarrow \overrightarrow{stms},
vjp_{stms}(stms),\overleftarrow{stms}$.  }

For example, the forward sweep of a multiplication
statement $\kw{let} \ x \ = \ a \ * \ b$ is the
statement itself. Re-executing it brings into scope
the (value of) variable $x$ which may be needed
in the return sweep of a following statement, e.g.,
$\kw{let} \ y \ = \ x \ * \ a$, which
would require an update to the adjoint of $a$ according
to the rewrite rule of~\cref{eq:rewrite-adj}:
$\overline{a} \ \mathrel{+}= \ \frac{\partial{(x * a)}}{\partial{a}}*\overline{y}$,
which results in $\overline{a} \ \mathrel{+}= \ x * \overline{y}$.

The rule for a loop statement is more complex.
Figure~\ref{fig:rules-loop} shows a loop that
exhibits a loop-variant variable $x$---whose value
changes through the loop---which is initialized
with $x_0$ just before the loop is started, and whose
value for the next iteration is bound to the
result of the loop body. The result of the entire
loop is bound to variable $y$. The rule assumes
that the shape of $x$ does not change throughout
the loop.

The forward sweep $\overrightarrow{loop}$ is the
original loop, except that its body is modified to checkpoint
into array $xs$ the value of $x$ at the entry of each
iteration.
As such $xs$ is also declared as loop variant
and initialized to $xs_0$, which, in turn, is
allocated (by $\kw{scratch}$) just before the
loop statement. Note that {\em only the
loops of the current scope are checkpointed};
an inner-nested loop would be re-executed
but not checkpointed.

The return sweep consists mainly of a loop that iterates
with $i$ from $n-1$ down to $0$:
the first statement re-installs the value of $x$
        for the current iteration from the checkpoint
        (i.e., $xs[i]$).
The remaining statements, $\overleftrightarrow{stms}$,
        are those generated by $vjp_{body}$ from the original
        loop body, including its forward sweep.
        As $\overleftrightarrow{stms}$ may use $x$, re-installment
        is necessary. Re-execution of
        the forward sweep brings into scope all variables
        that might possibly be used by the return sweep
        of the body.

The result of a reversed iteration is the adjoint of
        the original result $\overline{r}$, together with
        the (updated) adjoints of the free variables used
        in the loop $\overline{fvs_{bdy}'}$. These are 
        declared as variant through the loop
        ($\overline{fvs_{bdy}}$), such that the updates of
        all iterations are recorded. $\overline{fvs_{bdy_0}}$ is
        the adjoint of the free variables prior to entering the loop;
        if a free variable was used in a prior statement of
        the same scope (and hence has an existing potentially non-zero adjoint)
        its adjoint is simply retrieved. Otherwise, its adjoint is initialized
        to a zero element of the appropriate type and shape.

The final statement of the return sweep is
        $\overleftarrow{stm_{x_0}}$; it semantically
        updates the adjoint of the loop-variant initializer
        $\overline{x_0}$ with the (adjoint) result $\overline{x'}$
        of the return sweep loop---the code is generated
        by $vjp_{stm}(\kw{let} \ x'  =  x_0)$.\footnote{
          This is because the first (implicit) statement of the
          original loop is $\kw{let} \ x \ = \ x_0$, hence it
          is differentiated last on the return sweep. Moreover,
          the loop returns the (updated) adjoint of its variant
          variable $x$, which is bound/saved in $\overline{x'}$, and 
          is thus used to update the adjoint of $x_0$. 
        }

        Sequential loops are the only construct that require
        iteration checkpointing: \emph{parallel constructs do not}
        (e.g., because map iterations are independent by definition). 


\enlargethispage{\baselineskip}

For completeness, we conclude by treating array indexing.
When generating code for the
return sweep of a statement $\kw{let}~y~=~a[i]$, we must update
$\overline{a[i]}$ with the ``contribution'' of $\overline{y}$.
This leads us to the following rule:
\begin{equation*}
  \label{eq:vjp-index}
  vjp_{stm}(\kw{let}~y~=~a[i]) \Rightarrow (\kw{let}~y~=~a[i], \kw{let}~\overline{y}~=~\kw{upd}~i~\overline{y}~\overline{a})
\end{equation*}

The \kw{upd} construct merits further elaboration.
\emph{Semantically}, $\kw{upd}~i~v~a$ returns $a$ but with the value
at index $i$ changed to be $v+a[i]$.  \emph{Operationally}, the array
$a$ is directly modified in-place.  To preserve purely functional
semantics, we require that the ``old'' $a$ and its aliases are never
accessed again, similar to the in-place updates of \cref{sec:language}.

\subsection{Demonstrating Loops by Example \& Tradeoff}
\label{subsec:main-tradeoff}

\begin{figure}
\begin{subfigure}{0.55\linewidth}
\begin{lstlisting}[mathescape,basicstyle=\ttfamily\scriptsize,label={fig:loop-nest-orig},caption={Original Loop}]
$stms_{out}^{bef}$
let y'' =
  loop (y)=(y$_0$)
  for i = $0 \ldots m^k-1$ do
    $stms_{loop}$
    in y'
$stms_{out}^{after}$
\end{lstlisting}\vspace{-2ex}
\begin{lstlisting}[mathescape,basicstyle=\ttfamily\scriptsize,label={fig:loop-nest-strip},caption={Original loop\newline$\mbox{ }$\hspace{10ex}Stripmined $k\times$}]
$stms_{out}^{bef}$
let y'' =
  loop (y$_1$)=(y$_0$)
  for i$_1$ = $0 \ldots m-1$ do
    $\ldots$
    loop(y$_k$)=(y$_{k-1}$)
    for i$_k$ = $0 \ldots m-1$ do
      let y = y$_k$
      let i = i$_1$*$m^{k-1}$+$\ldots$+i$_k$
      $stms_{loop}$
      in y'
$stms_{out}^{after}$
\end{lstlisting}
\end{subfigure}
%
%
\begin{subfigure}{0.42\linewidth}
\begin{lstlisting}[mathescape,basicstyle=\ttfamily\scriptsize,label={fig:after-rev-ad},caption={Reverse AD\newline$\mbox{ }$Result on Original Loop}]
$\overrightarrow{stms_{out}^{bef}}$
let ys$_0$ = scratch($m^k$,
            sizeOf(y$_0$))
let (y'', ys) =
  loop (y,ys)=(y$_0$,ys$_0$)
  for i = $0 \ldots m^k-1$ do
    let ys[i] = y
    stms$_{loop}$
    in (y', ys)
$\overrightarrow{stms_{out}^{after}}$
$\overleftarrow{stms_{out}^{after}}$
let ($\overline{y'''}$, $\ldots$) =
  loop ($\overline{y}$,$\ldots$)=($\overline{y''}$,$\ldots$)
  for i = $m^k-1\ldots 0$ do
    let y = ys[i]
    $\overrightarrow{stms_{loop}}$
    $\overleftarrow{stms_{loop}}$
    in ($\overline{y'}$, $\ldots$)
$vjp_{smt}(\kw{let} \ y''' \ = \ y_0)$
$\overleftarrow{stms_{out}^{bef}}$
\end{lstlisting}
\end{subfigure}\vspace{-1ex}
\caption{ The figure shows the result of the reverse-AD transformation 
          (\ref{fig:after-rev-ad}) applied to code containing one loop 
          (\ref{fig:loop-nest-orig}). One can notice that the original-loop
          statements are re-executed twice ($stms_{loop}$ and
          $\vec{stms_{loop}}$).
        }
\label{fig:loop-eg}
\end{figure}

\cref{fig:loop-eg}, \cref{fig:after-rev-ad}, demonstrates the result
of the $vjp_{stms}$ transformation for the source code shown in
\cref{fig:loop-nest-orig}, which consists of a loop surrounded by
statements in the same scope ($stms_{out}^{bef}$,
$stms_{out}^{after}$).  With our strategy, the original loop is
executed twice: once on the forward sweep when it is checkpointed
($stms_{loop}$), and a second time on the return sweep, where the
forward sweep of the loop body ($\overrightarrow{stms_{loop}}$)
is responsible to bring into scope all the (original) variables
computed in the corresponding iteration. If the loop body consists
of scalar computations, re-computation is cheap. If the loop body
consists of nested recurrences, e.g., other loops, then the
re-computation overhead can be optimized, as before,
by creating perfect loop nests {\em and flattening them}. 

The latter is necessary because otherwise, the need of
checkpointing will keep them alive, and comes at the cost
of (asymptotically) increasing the memory footprint.
The time-space tradeoff~\cite{divide-and-conq} is
demonstrated in~\cref{fig:loop-nest-strip}:
applying the reverse AD to the strip-mined loop ($k = log_m m^k$ times)
increases the re-execution factor from $2\times$ to $(k+2)\times$,
but requires only $m k\times$ more memory than the original
program---i.e., the checkpointing of each of the $k$
strip-mined loops stores $m$ versions of the loop-variant
variable $y$; when $m$ is constant, this results in logarithmic
space and time overhead.

However, since GPUs have relatively small caches, we have
found this tradeoff useful only when AD runs out of memory.
In this sense, we allow the user to annotate the loops
with a constant stripmining factor, which is applied
automatically before the application of reverse AD.

\section{Rewrite Rules for Parallel Constructs}

Having defined the transformation across scopes, 
this section focuses on the mapping of individual parallel
constructs within a scope, though we allow arbitrary nesting
of them and loops together.
We organize the discussion in free-writing style fashion that focuses
on the reasoning that led to the derivation of the rules rather than
presenting them in a formal, but dry and verbose notation.  Some of
these rules resemble some that have been published
previously~\cite{10.1145/3471873.3472975}, but we developed them
independently~\cite{ad-lecture-slides}.

\subsection{Reduce and Multi-Reduce}

We recall that the semantics of reduce, for an arbitrary binary
associative operator $\odot$ and its neutral element $e_\odot$ is:\vspace{-1ex}
\begin{lstlisting}[mathescape] 
reduce $\odot$ $e_\odot$ [$a_0, a_1, \ldots, a_{n-1}$] $\equiv$ $a_0 \odot a_1 \odot \ldots \odot a_{n-1}$
\end{lstlisting}\vspace{-1ex}

If the result of \lstinline{reduce} is let-bound to variable
$y$, we can reason more easily about the contribution
of the reduce statement to the adjoint of $a_i$ if we group
terms as:\vspace{-1ex}
\begin{lstlisting}[mathescape] 
let $y$ = $(a_0 \odot \ldots \odot a_{i-1}) \ \odot \ a_i \ \odot \ (a_{i+1} \odot \ldots \odot a_{n-1})$
\end{lstlisting}\vspace{-1ex}
If we would know, for every $i$, the terms
$l_i = a_0 \odot \ldots \odot a_{i-1}$ and 
$r_i = a_{i+1} \odot \ldots \odot a_{n-1}$, 
we could directly apply the main rule for reverse AD
given in \cref{eq:rewrite-adj}, which results in:
 \[ \overline{a_i} \ \ \ \overline{+}= \ \ \frac{\partial{(l_i \ \odot \ a_i \ \odot \ r_i)}}{\partial{a_i}} \ \overline{y} \]
\noindent where $l_i$ and $r_i$ are considered constants and
$\overline{+}$ denotes a potentially vectorized
addition that matches the datatype. 
The code for the right-hand side (RHS) can be generically 
generated as a function $f$ that can be {\em mapped} to all $a_i$, $l_i$, $r_i$:
\[ f \ \leftarrow \ vjp_{\lambda}( \ \overline{y}, \ \lambda (l_i, \ a_i, \ r_i) \rightarrow l_i \ \odot \ a_i \ \odot \ r_i \ )\]
except that the adjoints of $l_i$ and $r_i$ are not returned by $f$.

Finally, all the $l_i$ and $r_i$ values can be computed
by a forward and reverse exclusive scan (prefix sum),
respectively. Essentially, the forward sweep is the
reduce statement. Denoting by $as = [a_0, a_1, \ldots, a_{n-1}]$,
and assuming for simplicity that $\odot$ has no free variables,
the return sweep is:\vspace{-1ex}
\begin{lstlisting}[mathescape]
let $ls$ = scan$^{exc}$ $\odot$ $e_\odot$ $as$
let $rs$ = reverse $as$ |>
        scan$^{exc}$(\x y -> y $\odot$ x) e$_\odot$ |> reverse
let $\overline{as}$ $\overline{+}$= map f $ls$ $as$ $rs$
\end{lstlisting}\vspace{-1ex}

Essentially, the reverse AD code for an arbitrary reduce
requires a map, two scan\footnote{
{\tt scan$^{exc} \odot~e_\odot$ [$a_0, \ldots, a_{n-1}$] $\equiv$ [$e_\odot, a_0, a_0 \odot a_1,\ldots, a_0 \odot \ldots \odot a_{n-2}]$}
} and two reverse operations, which
preserve the work-depth parallel asymptotics of the original 
program. In practice, this is quite expensive,\footnote{
In theory, the adjoint code should not be more than
$4\times$ slower than the original,
but our general rule requires at least $5$ global memory
accesses per array element ($2$ for each scan), while the 
original reduce requires only one.
} but luckily
most standard operators, admit more efficient translations:

\subsubsection{Special Cases of Reduce Operators}
The treatment of plus, min, max is known~\cite{four-spec-cases},
but we present a more efficient rule for multiplication.
We recount them here for completeness and because they
are used for reduce-by-index, which, to our knowledge,
has not been covered before.

When the reduce operator is (vectorized) \textbf{\em plus}, we have
$\frac{\partial{(l_i \ + \ a_i \ + \ r_i)}}{\partial{a_i}} \ \overline{y} \ = \ \overline{y}$,
hence the return sweep becomes\\
$\kw{let} \ \overline{as} \ \ \overline{\tt+}\text{= } \ \overline{y}$,
which is also derived automatically by the simplification engine from the general rule.\smallskip

When the reduction operator is \textbf{\em multiplication},
we have $\frac{\partial{(l_i \ * \ a_i \ * \ r_i)}}{\partial{a_i}} \ \overline{y} \ = \ l_i * r_i * \overline{y}$. We discriminate three cases here: 
\begin{itemize}
\item if all $as$' elements are nonzeros, then 
        $l_i * r_i \ = \frac{y}{a_i}$ and $y \neq 0$, hence we
        update each element:
        $\overline{a_i} \ \ {\tt+}\text{= } \ \frac{y}{a_i} \overline{y}$,
        
\item if exactly one element at index $i_0$ is zero, then
        $l_i * r_i$ is zero for all other elements, and
        $\ \ \overline{a_{i_0}} \ \ {\tt+}\text{= } \ y * \overline{y} $
\item otherwise, $\forall i, l_i*r_i \equiv 0$, and $\overline{as}$
        remains unchanged.
\end{itemize}
The forward sweep is modified to compute the number of 
    zeros in $as$ and the product of non-zero
    elements (by a map-reduce operation), followed
    by setting the reduced result $y$ accordingly.
The return sweep computes the contributions
by a parallel map and updates adjoints as discussed
before.\smallskip

Finally, when the reduction operator is \textbf{\em min} (or
\textbf{\em max}), then only the adjoint of (one of) the minimal array
element should receive the contribution of $y$, because it is the only
one that was used 
to compute the result. As such the forward sweep is modified to
compute the minimal element together with its (first) index $i_y$ (the
common ``argmin'' operation), which can be done with a parallel
reduction.  The forward sweep is then
$\kw{let}\ \ (y,i_{y}) = \textrm{argmin}\ as$, and the return sweep
could be
$\kw{let} \ \ \overline{as}[i_y] \ \ \overline{\text{+}}= \ \
\overline{y}$.

However, at this point we add a compiler improvement that
exploits sparsity: For example, if $\overline{as}$
has not been created yet, then instead of inserting the update
statement, we (statically) record in $vjp$'s environment that
$\overline{as}$ is sparse (currently one non-zero element).
Further operators such as \texttt{\kw{reduce} min}, may refine
the sparse structure of $\overline{as}$ by adding more points
to it. If $as$ is created by a $\kw{map}$ operation, and
$\overline{as}$ is still sparse, then we (statically)
replace the adjoint code of the $\kw{map}$ with (only) the adjoint
code of the iterations corresponding to the non-zeros in
$\overline{as}$---because the other iterations have no effect.


\subsubsection{Reduce-by-Index}
Reduce-by-index~\cite{histo-sc20}, a.k.a. multi-reduce,
essentially generalizes a histogram computation by allowing
the values from an array ($as$) that fall into the same bin
(index from {\tt inds}) to be reduced with an arbitrary
associative and commutative operator $\odot$ having neutral 
element $e_\odot$, where the number of bins $m$ is typically
assumed smaller than that of index-value pairs $n$, i.e.,
%
%
\begin{lstlisting}[mathescape]
let hs = reduce_by_index ($\odot$) $e_\odot$ inds $as$
\end{lstlisting}\vspace{-1ex}
has the semantics:\vspace{-1ex} 
\begin{lstlisting}[mathescape]
loop hs = replicate m $e_\odot$
for i = 0 $\ldots$ n-1 do
  let hs[ inds[i] ] $\odot$= $as$[i] in hs
\end{lstlisting}\vspace{-1ex} 
A similar reasoning as for the arbitrary \lstinline{reduce}
suggests that two scans need to be applied
to each subset of elements that fall in the same bin, 
a.k.a., multi-scan, in order to compute the $l_i$ and
$r_i$ terms for every $i$.  Then the contributions to
the adjoint of $as$ of a reduce-by-index statement are
computed as before by \lstinline{map f ls as rs}.
Assuming a constant key size, the multi-scan can be
implemented within the right work-depth asymptotic by
(radix) sorting $as$ according to the corresponding
bins, and then by applying irregular-segmented scans
(forward and reverse) on the result. Work is in
progress to implement this case. 
We have implemented however the special-case operators
discussed for reduce. Essentially, the forward sweep
consists of the reduce-by-index statement, but enhanced
with the extended operators, and the return sweep
is similar to reduce, except that in the update formula
of the adjoint $\overline{as}\text{[i]}$, $\overline{y}$
is replaced with {\tt$\overline{\text{hs}}$[inds[i]]}.




\subsection{Scan or Prefix Sum}
\label{scan-rule}

An inclusive scan~\cite{segScan} computes all prefixes of an array by means of
an associative operator $\odot$ with neutral element $e_\odot$:\vspace{-1ex}
\begin{lstlisting}[mathescape] 
scan $\odot$ $e_\odot$[$a_0, \ldots, a_{n-1}$] $\equiv$ [$a_0, a_0 \odot a_1,\ldots, a_0 \odot \ldots \odot a_{n-1}]$
\end{lstlisting}\vspace{-1ex}
While the derivation of (multi-) reduce builds on a functional-like
high-level reasoning, in scan's case, we found it easier to reason
in an imperative, low-level fashion. For simplicity we assume first
that $\odot$ operates on reals, and generalize later:\vspace{-1ex}
\begin{lstlisting}[mathescape,basicstyle=\ttfamily\scriptsize]
let $rs$[0] = $as$[0]
let $ys$ = loop ($rs$) = ($rs$) for i = 1 $\ldots$ n$-$1 do
            let $rs$[i] = $rs$[i-1] $\odot$ $as$[i] in $rs$ 
\end{lstlisting}\vspace{-1ex}
The loop above that implements scan, writes each element of the
result array {\tt rs} exactly once. To generate its return sweep,
we can reason that we can fully unroll the loop, then
apply the main rewrite-rule from~\cref{eq:rewrite-adj} to each
statement and finally gather them back into the loop below:\vspace{-1ex}
\begin{lstlisting}[mathescape,basicstyle=\ttfamily\scriptsize]
let ($\overline{rs'},\overline{as}$) = loop ($\overline{rs},\overline{as}$) = (copy $\overline{ys},\overline{as}$) for i = n-1..1 do
    let $\overline{rs}$[i-1] += $\frac{\partial \ (rs[i-1] \ \odot \ as[i])}{\partial \ rs[i-1]} * ~ \overline{rs}[i]$
    let $\overline{as}$[i] += $\frac{\partial \ (rs[i-1] \ \odot \ as[i])}{\partial \ as[i]} * ~ \overline{rs}[i]$    in ($\overline{rs}$, $\overline{as}$)
let $\overline{as}$[0] += $\overline{rs'}$[0]
\end{lstlisting}\vspace{-1ex}
Simple dependence analysis, for example based on direction
vectors, shows that the loop can be safely distributed across
its two statements, since they are not in a dependency cycle:\vspace{-1ex}
\begin{lstlisting}[mathescape,basicstyle=\ttfamily\scriptsize]
let ($\overline{rs'}$) = loop ($\overline{rs}$) = (copy $\overline{ys}$) for i = n-1$\ldots$1 do
             let $\overline{rs}$[i-1] += $\frac{\partial \ (rs[i-1] \ \odot \ as[i])}{\partial \ rs[i-1]} ~*~ \overline{rs}[i]$ in $\overline{rs}$

let ($\overline{as}$) = loop ($\overline{as}$) = ($\overline{as}$) for i = n-1$\ldots$0 do
  	         let el = if i==0 then $\overline{rs'}$[0]
  			              else $\frac{\partial \ (rs[i-1] \ \odot \ as[i])}{\partial \ as[i]} ~*~ \overline{rs'}[i]$
  	         let $\overline{as}$[i] += el in $\overline{as}$
\end{lstlisting}\vspace{-1ex}

The second loop (computing $\overline{as}$) is essentially a 
\lstinline{map} once we know the values of $\overline{rs'}$.  
Denoting by
$c_{n-1} = 1$ and 
$c_{i} = \frac{\partial \ (rs_{i} \ \odot \ as_{i+1})}{\partial \ rs_{i}}$, 
the first loop (computing $\overline{rs}$) is
a backward linear recurrence of the form\vspace{-1ex}
\begin{lstlisting}[mathescape]
$\overline{rs}_{n-1} = \overline{ys}_{n-1},$    $\overline{rs}_{i} = \overline{ys}_{i} ~+~ c_{i}\cdot\overline{rs}_{i+1},~~~i=n-2 ... 0$
\end{lstlisting}\vspace{-1ex}
where $\overline{ys}$ is the adjoint of the result of the original
loop statement, which is known before the reversed loop is entered.

Such a recurrence is known to be solved with a scan whose operator
is linear-function composition~\cite{Blelloch1990PrefixSA}. 
To summarize: the forward sweep is the original scan. The return
sweep consists of 
(1) the \lstinline{map} that computes the $c_i$ values,
(2) the \lstinline{scan} that computes the backward linear recurrence, and
(3) the \lstinline{map} that computes that updates $\overline{as}$:\vspace{-1ex}
\begin{lstlisting}[mathescape,basicstyle=\ttfamily\scriptsize]
let (ds, cs) =
    map ( \i -> if (i == n-1) then (0, 1)
               else ($\overline{ys}[i]$, $\frac{\partial \ (rs[i] \ \odot \ as[i+1])}{\partial \ rs[i]}$) ) [0..n-1]
let lin$_o$ (d1,c1) (d2,c2) = (d2 + c2 $\cdot$ d1, c2 $\times$ c1)
let $\overline{rs}$ = scan lin$_o$ (0,1) (reverse ds) (reverse cs)
      |> map (\ $d_i$ $c_i$ -> $d_i$ + $c_i \cdot \overline{ys}$[n-1]) |> reverse

let $\overline{as}$ $\overline{+}$= map (\ (i, a$_i$) -> if i==0 then $\overline{rs}$[0]
                           else $\frac{\partial \ (rs[i-1] \ \odot \ a_i)}{\partial \ a_i} \cdot \overline{rs}[i]$
              ) [0..n-1] as
\end{lstlisting}\vspace{-1ex}

For generalization, we observe that any element type can be
linearized to a vector, say of size $d$.  Then we observe that
\begin{itemize}
\item a term like $\frac{\partial \ (rs[i-1], a_i)}{\partial \ a_i}$
        corresponds to the Jacobian of the function 
        $\odot_{right}(y) \ = \ rs[i-1] \odot y$ at point $a_i$,
        denoted by $J_{\odot_{right}}(a_i) \ \in \ \mathbb{R}^{d\times d}$,
\item a term like $\frac{\partial \ (rs[i-1], a_i)}{\partial \ a_i}  \cdot  \overline{rs}[i] \ \in \ \mathbb{R}^d$
        applies the Jacobian to a vector and corresponds to
        $vjp_{\lambda}(\overline{rs}[i], \odot_{right})$.
\end{itemize}
It follows that in the rewrite rule above, $\cdot$ and $\times$
are generalized from scalar multiplication to matrix-vector and
matrix-matrix multiplication, respectively, and $+$ to vector
addition. In particular, the $lin_o$ operator of scan, has
neutral element $(0\in \mathbb{R}^d, I \in \mathbb{R}^{d\times d})$,
where $I$ is the identity matrix, and $lin_o$ computes, among
others, one matrix multiplication.
If $d$ is a constant, e.g., tuples of scalars, the work-depth
asymptotic of the original program is preserved, otherwise, 
this rule provides no such guarantee. We have implemented the
case when scan's element type is one scalar. We also support
vectorized such operators by turning them beforehand to a 
regular-segmented scan, by the rule:\vspace{-1ex}
\begin{lstlisting}[mathescape]
scan (map ($\odot$)) $\overline{e_\odot}$ xs $\Rightarrow$
transpose xs |> map (scan $\odot$ $e_\odot$) |> transpose
\end{lstlisting}\vspace{-1ex}
We treat separately the case of scan with (vectorized)
plus operator, in which the contributions to be accumulated
by $\overline{as}$ are given by
$\kw{scan} \ (\overline{+}) \ \overline{0} \ (\kw{reverse} \ \overline{ys}) \ \pipe \ \kw{reverse}$.

\subsection{Parallel Scatter}
\label{scatter-rule}

The statement \lstinline{let ys = scatter xs is vs} produces an array
{\tt ys} by updating in-place the array {\tt xs} (which is consumed)
at the $m$ indices specified in array {\tt is} with corresponding
values taken from {\tt vs}. Scatter has constant depth, and work
proportional with $m$, the size of array {\tt vs}, i.e., it does not
depends of the length $n$ of the updated array {\tt ys}.  Our rule
assumes that {\tt is} contains no duplicates, i.e., idempotent updates
are currently not supported.

Denoting by \kw{gather} the operation that reads from a 
support array elements at a given subset of indices, i.e.,\vspace{-0.5ex}
\begin{lstlisting}[mathescape]
let gather arr inds = map (\i -> xs[i]) inds
\end{lstlisting}\vspace{-0.5ex}
the forward sweep saves (in {\tt xs$_{saved}$})
the elements of $xs$ that are about to be overwritten
prior to performing the update:
\begin{lstlisting}[mathescape]
let xs$_{saved}$ = gather xs is
let ys = scatter xs is vs
\end{lstlisting}
Since the original statement performs
{\tt ys[ is[i] ] = vs[i]} for any $i\in\mbox{\tt is}$,
we have by~\cref{eq:rewrite-adj}  
{\tt $\overline{vs}$[i] += $\overline{ys}$[is[i]]}.
As such, the return sweep (1) first updates the
adjoint of {\tt vs}, then (2) creates the adjoint
of {\tt xs} by zeroing out the elements from $\overline{ys}$
that were subject to the scattered update---because those
adjoints correspond to elements that were never in {\tt xs},
and finally, (3) restores {\tt xs} to its state before the
update:\vspace{-0.5ex}
\begin{lstlisting}[mathescape]
let $\overline{vs}$ $\overline{+}$= gather is $\overline{ys}$
let $\overline{xs}$ = scatter $\overline{ys}$ is (replicate m 0)
let xs = scatter ys is xs$_{saved}$
\end{lstlisting}\vspace{-0.5ex}

Note that both forward and return sweep preserve the
original work-depth asymptotics, because all operations have
work proportional to $m$ (not $n$). This would not hold
if, e.g., the forward sweep would make a copy of the whole
{\tt xs}.

%
%

\subsection{Map}
\label{sec:map}

A \kw{map} applies a lambda function to each element of an
array, producing an array of same length:
\[
\kw{let}~xs~=~\kw{map}~(\lambda a \rightarrow stms~\kw{in}~x)~as
\]
If the lambda has no free variables, the return sweep is:
\[
  \begin{array}{r@{}c@{}l@{}l}
    \kw{let}~\overline{as}
    &~=~
    &\kw{map}~
    &(\lambda(a,\overline{a},\overline{x}) \rightarrow \overrightarrow{stms}~\overleftarrow{stms}~\kw{in}~\overline{a_0} + \overline{a})\\
    &&&as~\overline{as}~\overline{xs}
  \end{array}
\]
where $\overrightarrow{stms}$ and $\overleftarrow{stms}$ denote the
forward and reverse statements of lambda's body, and $\overline{a}$ is
shadowed by $\overleftarrow{stms}$.


A naive way of handling free variables is to turn them into bound
variables.  E.g. converting
$\kw{map}~(\lambda i \rightarrow as[i])~is$ into
$\kw{map}~(\lambda (i,as') \rightarrow
as'[i])~is~(\kw{replicate}~n~as)$ where $n$ is the size of
$is$.  This is fine for scalars, but asymptotically inefficient for
arrays that are only partially used, as here, as the
adjoint will be mostly zeroes.

In an impure language, we could update the adjoint of a free array
variable $as[i]$ with an operation $\overline{as}[i]\texttt{+=}v$,
implemented with atomics or locks in the parallel case.  In our pure
setting, we instead introduce \emph{accumulators}.  An array can be
``temporarily'' turned into an accumulator with
\kw{withacc}\footnote{For simplicity we treat only single-dimensional
  arrays in this section, but the idea also works in the
  multi-dimensional case.  This type for \kw{withacc} allows only a
  single result corresponding to the array being updated.  In
  practice, we also need to be able to return an arbitrary secondary
  result.}:
\[
  \kw{withacc} : [d]\alpha \rightarrow (\kw{acc}(\alpha) \rightarrow \kw{acc}(\alpha)) \rightarrow [d]\alpha
\]
Intuitively we view an accumulator as a ``write-only'' view of an
array.  \emph{Semantically}, accumulators are lists of index/value
pairs, each denoting an update of an array.  When we use \kw{upd} on
an accumulator, we prepend index/value pair to this list, returning a
new accumulator. \emph{Operationally}, \kw{upd} on an accumulator is
implemented by immediately updating the underlying array, not by
actually maintaining a list of updates in memory (although such an
implementation would be semantically valid).  The purpose of
accumulators is to allow the compiler to continue to reason purely
functionally, in particular that all data dependencies are explicit,
while allowing efficient code generation based on in-place updates.
Our accumulators are similar to generalized
reductions~\cite{genred-crummey} or the accumulation effects in
Dex~\cite{10.1145/3473593} and have the same underlying motivation.
The main difference is that in Dex, they are an effect, which requires
every part of the compiler to be effect-aware.

Free array-typed variables in \kw{map} are thus turned into
accumulators while generating return sweep code for the \kw{map},
during which we can perform the updates directly.  We allow implicit
conversion between accumulators and arrays of accumulators, as this
allows us to directly \kw{map} them.  E.g.
\[\kw{let}~xs~=~\kw{map}~(\lambda i \rightarrow as[i])~is\]
results in the return sweep code
\[
  \begin{array}{l}
    \kw{let}~\overline{as}~=~
    \kw{withacc}~\overline{as}~(\lambda \overline{as}_{\textrm{acc}} \rightarrow \\
    \quad \kw{map}~
    (\lambda (i,\overline{x},\overline{as})\rightarrow\kw{upd}~i~\overline{x}~\overline{as})~
    is~\overline{xs}~\overline{as}_{\textrm{acc}})
  \end{array}
\]
where we treat $\overline{as}_{\textrm{acc}}$ as an array of
accumulators when passed to \kw{map}, and treat the result of the
\kw{map} as a single accumulator.  This is efficient because
accumulators have no run-time representation, and saves us from
tedious boilerplate.

During the lifetime of the accumulator, the underlying array may not
be used---this prevents observation of intermediate state.  These
rules can be encoded in a linear type system and mechanically checked,
which we do in our implementation, but exclude from the paper for
simplicity.

Accumulators are sufficient to express the adjoint computation inside
maps because (1) any read from an array $a[i]$ is turned into an
accumulation on $\overline{a}[i]$ as discussed at the end of \cref{transf-scopes},
and (2) the only place on the return sweep where $\overline{a}$ can be
read outside an accumulation statement is the definition of $a$, which
by definition is the last use of $\overline{a}$, hence it is safe to
turn it back into a normal variable there.

\section{Implementation and Optimizations}

We have implemented the reverse AD transformation as a pass
in a copy of the publicly available Futhark compiler.\footnote{
\url{https://github.com/diku-dk/futhark}
}
The presented transformation rules were tuned to preserve
fusion opportunities, both with constructs from the
statement's sweep, and across statements. 
\cref{subsec:false-deps} discusses several omitted issues,
namely how to optimize checkpointing for the arrays that
are constructed by in-place updates inside loop nests, and
how to support while loops, and \cref{subsec:limits}
presents the limitations of our implementation.

Since accumulators were not supported in the original language,
we have implemented them throughout the compiler---for the GPU
backends, they ultimately boil down to atomic updates, such as
{\tt atomicAdd} in CUDA.
Accumulators however, often result in suboptimal performance,
because they access memory in uncoalesced fashion and are
subject to conflicts, i.e., threads simultaneously accessing
the same location.
In this sense, \cref{subsec:opt-mat-mul} presents optimizations
aimed at turning accumulators into more specialized constructs
(e.g., map-reduce) that can be better optimized.

\subsection{Optimizing Accumulators}
\label{subsec:opt-mat-mul}

We demonstrate our optimizations of accumulator on the
matrix-matrix multiplication. The code below assumes
$as \in \mathbb{R}^{m\times q}$ and $bs\in \mathbb{R}^{q\times n}$
and computes $cs \in \mathbb{R}^{m\times n}$ by taking
the dot product of each (pair of) row of $as$ and column of $bs$:
\begin{lstlisting}[mathescape,basicstyle=\ttfamily\scriptsize]
let xs = map (\i -> map (\j -> sum (map (*) as[i,:] bs[:,j])
             ) [0$\ldots$n-1]) [0$\ldots$m-1]
\end{lstlisting}

Differentiating the code above results in the return sweep:
\begin{lstlisting}[mathescape,basicstyle=\ttfamily\scriptsize]
let ($\overline{as}, \overline{bs}$) = withacc($\overline{as} \ \overline{bs}$ \$\overline{as} \ \overline{bs}$ ->
  map (\i $\overline{as}$ $\overline{bs}$ -> map (\j $\overline{as}$ $\overline{bs}$ -> map (\k $\overline{as}$ $\overline{bs}$ ->
        -- $\overline{as}$[i,k] += bs[k,j]*$\overline{cs}$[i,j]
        let $\overline{as}$ = upd (i,k) (bs[k,j]*$\overline{cs}$[i,j]) $\overline{as}$
        -- $\overline{bs}$[k,j] += as[i,k]*$\overline{cs}$[i,j]
        let $\overline{bs}$ = upd (k,j) (as[i,k]*$\overline{cs}$[i,j]) $\overline{bs}$ in ($\overline{as},\overline{bs}$)
      ) [0$\ldots$q-1] $\overline{as}$ $\overline{bs}$) [0$\ldots$n-1] $\overline{as}$ $\overline{bs}$) [0$\ldots$m-1] $\overline{as}$ $\overline{bs}$
\end{lstlisting}
which is not efficient, because (temporal) locality is sub-utilized.
In this sense, we have designed and implemented a pass aimed at
turning (common cases of) accumulator accesses into reductions.
We present the analysis at an intuitive level: The analysis searches
for the first accumulator directly nested
in a perfect map-nest and checks whether its indices are invariant
across any of the parallel dimensions. In our case $as$ is
accumulated on indices $[i,k]$ that are both invariant to the
outermost parallel index $j$. In such a case,
the map-nest is split into two: the code on which the 
accumulated statement depends, and the code without the
accumulator statement,\footnote{
The optimization fires only if the number of redundant access to
global memory introduced by splitting the map nest is less than two.
} which is simplified and treated recursively.
The map-nest encapsulating the accumulation is
reorganized such that the invariant dimension ($j$) is moved
innermost.\footnote{
  It is always safe to interchange parallel loops inwards.
}
The accumulated statement is taken out of this
innermost map, which is modified to produce (only) the
accumulated values, whose sum (\lstinline{reduce (+) 0})
is re-written to be the value accumulated by $\overline{as}$:
\begin{lstlisting}[mathescape,basicstyle=\ttfamily\scriptsize]
let ($\overline{as}$) = withacc($\overline{as}$ \$\overline{as}$ ->
  map (\i $\overline{as}$ -> map (\k $\overline{as}$ ->
        let s = sum (map (*) bs[:,j] $\overline{cs}$[i,:])
        in  upd (i,k) s $\overline{as}$ -- $\overline{as}$[i,k] += bs[k,j]*$\overline{cs}$[i,j]
      ) [0$\ldots$q-1] $\overline{as}$) [0$\ldots$m-1] $\overline{as}$
let ($\overline{bs}$) = withacc($\overline{bs}$ \$\overline{bs}$ ->
  map (\k $\overline{bs}$ -> map (\j $\overline{bs}$ ->
        let s = sum (map (*) as[:,k] $\overline{cs}$[:,j])
        in  upd (k,j) s $\overline{bs}$ -- $\overline{bs}$[i,k] += as[i,k]*$\overline{cs}$[i,j]
      ) [0$\ldots$n-1] $\overline{bs}$) [0$\ldots$q-1] $\overline{bs}$
\end{lstlisting}

One can notice that the code now essentially consists
(as expected) of two matrix-multiplication-like kernels.
These are optimized by a later pass that performs
block and register tiling whenever it finds an innermost
map-reduce whose input arrays are invariant to one of
the outer-parallel dimensions. We have extended this pass
(1) to support accumulators, (2) to keep track of the array
layout---i.e., transposed or not, (3) to copy from global to
shared memory in coalesced fashion for any layout, and 
(4) to exploit some of the parallelism of the innermost
dot product as inspired from~\cite{ari-mmm}.

This optimization is responsible for nearly a one-order-of-magnitude
speedup at application level for benchmarks dominated by
matrix multiplication such as GMM and LSTM, evaluated in
sections~\ref{sec:gmm}~and~\ref{sec:lstm}.

\subsection{Loop optimizations}
\label{subsec:false-deps}


As discussed in section~\ref{sec:tape-loops}, by default,
loop-variant variables are saved/checkpointed at the entry
of each iteration.
This technique does not preserve the work asymptotic
of the original program when a loop variant array is
modified in place. For example, the contrived dependent
loop below constructs an array of length $n$ in 
$O(n)$ work, but the checkpointing of the forward sweep
would require $O(n^2)$ work:
\begin{lstlisting}[mathescape]
loop (xs) = (xs_0) for i = 1$\ldots$n-1 do
  xs with [i,j] = ass[i,j] + xs[i-1,j]
\end{lstlisting}
One can observe however that iteration-level checkpointing
is not needed if the loop nest does not exhibit {\em any false
dependencies} ({\sc war}+{\sc waw}):\footnote{
 The absence of false dependencies, means that the loop
    has only true ({\sc raw}) dependencies or no dependencies
    at all.
}
since no value is ``lost'' through the loop nest, it is
enough to checkpoint {\tt xs} only
once at the entry to the outermost loop of the nest. 
Moreover, re-execution is safe because all the
over-writes are idempotent.  In this sense, we
allow the user to annotate such loop parameters
that are free of false dependencies:
we checkpoint them at the loop-nest entry
and restore this copy just before entering the
return sweep of that nest.   Essentially, 
we consider that false dependencies are properly named
as such, and the user should not rely on them, especially
in a data-parallel context.
Work in the automatic parallelization area can be used
to automatically check the safety of
such annotations, statically~\cite{SUIF},
dynamically~\cite{R-LRPD}, and
anywhere in between~\cite{CosPLDI}.

%

A second issue relates to while loops, on which we cannot perform
AD directly, because their statically-unknown iteration count hinders
the allocation of the checkpointing tape.
We provide two mechanisms to address this issue: first, the user
may annotate a while-loop with an iteration bound $n$. The while loop
is then transformed into an $n$-iteration for-loop that contains
a perfectly nested $\kw{if}$, which only executes the valid iterations
of the while loop. Alternatively, in the absence of such annotation,
the loop count can be computed dynamically by an inspector---i.e.,
loop slice that computes only the number of iterations of the loop.

\subsection{Current Implementation Limitations}
\label{subsec:limits}

We have already discussed limitations during the presentation
of individual constructs. The main remaining one is that
we do not currently support loop-variant parameters that
change their shape throughout the loop.
In principle this can be handled by dynamic re-allocations,
albeit this might prove expensive in a GPU setting.

Another shortcoming is that when an in-place update occurs inside
an $\kw{if-then-else}$, we currently save the target array at the
branch's entry: in principle, we should instead propagate the saved
element(s)---denoted $xs_{saved}$ in \cref{scatter-rule}---outside
branches, and back in to reach the return sweep.\footnote{Of course, if the in-place update 
occuring inside the $\kw{if}$ is protected by the annotation
discussed in \cref{subsec:false-deps}, then they do not
require any instrumentation.} By construction, this propagation
cannot escape the scope of the innermost-enclosing recurrence.

\section{Experimental Evaluation}

\subsection{AD-Bench: Sequential AD Performance}
\label{sec:adbench}

ADBench is a collection of benchmarks aimed at comparing different AD
tools~\cite{Srajer2018Abo}.  We have used the ADBench framework to
measure Futhark versions of the four ADBench problems.  We compile to
sequential CPU code and report the time taken to compute the full
Jacobian relative to the time taken to compute the objective function,
using the largest default dataset for each benchmark.  We compare
against Tapenade~\cite{Araya-Polo2004DFA} and manually differentiated
programs.  The results are shown in \cref{tab:adbench}.

\begin{table}
  \begin{tabular}{c|rrrrr}
    \textbf{Tool}
    & \multicolumn{1}{c}{\textbf{BA}}
    & \multicolumn{1}{c}{\textbf{D-LSTM}}
    & \multicolumn{1}{c}{\textbf{GMM}}
    & \multicolumn{2}{c}{\textbf{HAND}} \\
    &&&& \textbf{Comp.} & \textbf{Simple} \\ \hline
    \textbf{Futhark}  & $13.0\times$ & $3.2\times$ & $5.1\times$ & $49.8\times$ & $45.4\times$ \\
    \textbf{Tapenade} & $10.3\times$ & $4.5\times$ & $5.4\times$ & $3758.7\times$ & $59.2\times$ \\
    \textbf{Manual}   & $8.6\times$ & $6.2\times$ & $4.6\times$ & $4.6\times$ & $4.4\times$ \\
  \end{tabular}
  \caption{Time to compute the full Jacobian relative to time to
    compute the objective function.  Lower is better.}
  \label{tab:adbench}
\end{table}

Futhark does well, in particular managing to outperform Tapenade in
four out of five cases.  For the exception, BA, the bottleneck is
packing the result (which is a sparse Jacobian) in the CSR format
expected by the tooling, which is code that is not subject to AD.
The HAND benchmark has two variants: a ``simple'' one that computes
only the dense part of the Jacobian, and a ``complicated'' one that
also computes a sparse part.  Tapenade handles the latter
poorly.  Both Tapenade and Futhark perform poorly when compared to the
hand-written code.
Both BA and HAND produce sparse Jacobians where the sparsity structure
is known in advance, which is exploited by passing appropriate seed
vectors to \texttt{jvp}/\texttt{vjp}.

\subsection{Parallel Hardware and Methodology}

We benchmark on two different Linux systems: one with two AMD EPYC
7352 CPUs and an NVIDIA A100 GPU running CUDA 11.2, which we denote
\textbf{A100}, and another with two Intel E5-2650 CPUs and an NVIDIA
RTX 2080Ti GPU running CUDA 11.3, which we denote \textbf{2080Ti}. We
report mean runtime for 10 runs, which includes all overheads, except
transfering program input and result arrays between device and host.
We report the absolute runtime of the differentiated and primal
program, and the ``overhead'' of differentiation that corresponds to
the ratio between the two. In optimal AD, this ratio (counted in
number of operations) is supposed to be a small
constant~\cite{griewank2008evaluating}, hence the ratio serves as a
good measure of the efficiency of an AD implementation.
Futhark benchmarks using 32-bit floats were performed with
block and register tile sizes of 16 and 4, respectively; on 64-bit
floats the register tile size is 2.


\subsection{Compared to Enzyme}
\label{sec:enzyme}





RSBench and XSBench are CUDA implementations of Monte Carlo neutron
transport algorithms that are reported in the Enzyme
paper~\cite{10.1145/3458817.3476165}, and which we have ported to
Futhark in order to quantiatively compare our solution with Enzyme.
They each constitute a large \texttt{map} operation that contains
inner loops and control flow, as well as indirect indexing of arrays.
We benchmark on the \textbf{2080Ti} system, which features a GPU
similar to the one used to report the Enzyme results.  We use the
``small'' datasets.  The results (\cref{tab:rsbench-xsbench}) show the
overhead caused by our AD and the one reported in the Enzyme paper.
Our overhead is slightly smaller, although this may come down to
micro-optimisations.

\begin{table}
  \centering
  \begin{tabular}{l|cc|cc}
    & \multicolumn{2}{c|}{\textbf{Primal runtimes}} & \multicolumn{2}{c}{\textbf{AD overhead}} \\
    & \textbf{Original} & \textbf{Futhark} & \textbf{Futhark} & \textbf{Enzyme} \\\hline
    \textbf{RSBench} & $2.311s$ & $2.118s$ & $3.6\times$ & $4.2\times$ \\
    \textbf{XSBench} & $0.244s$ & $0.235s$ & $2.6\times$ & $3.2\times$
  \end{tabular}
  \caption{Results for RSBench and XSBench.  We report the absolute
    runtimes of the original program and the un-differentiated Futhark
    program, as well as the overhead of a single forward and return
    sweep of the reverse-mode differentiated program compared to the
    un-differentiated one.}
  \label{tab:rsbench-xsbench}
\end{table}

\subsection{Case Study 1: Dense $K$-means clustering}
\label{sec:dense-kmeans}

\begin{table}
  \centering
  \begin{tabular}{lc|rr|r}
                                                                       & & \multicolumn{2}{c|}{\textbf{Futhark}} & \\\
                                                                       & $(k,n,d)$ & \textbf{Manual} & \multicolumn{1}{c|}{\textbf{AD}} & \textbf{PyTorch} \\\hline
 \multirow{2}{*}{\rotatebox[origin=c]{90}{\scriptsize\textbf{A100}}}   & $(5,494019,35)$    & $9.3ms$ & $36.6ms$ & $44.9ms$ \\
                                                                       & $(1024,10000,256)$ & $9.9ms$ & $9.6ms$ & $11.2ms$ \\ \hline
 \multirow{2}{*}{\rotatebox[origin=c]{90}{\scriptsize\textbf{2080Ti}}} & $(5,494019,35)$    & $17.8ms$ & $72.6ms$ & $71.4ms$\\
                                                                       & $(1024,10000,256)$ & $18.8ms$ & $19.1ms$ & $22.6ms$ \\
  \end{tabular}
  \caption{Performance measurements for $k$-means clustering for two
    different workloads, implemented both manually in Futhark, with AD
    in Futhark, and with AD in PyTorch.}
  \label{tab:kmeans}
\end{table}

$k$-means clustering is an optimization problem where given $n$ points
$P$ in a $d$-dimensional space we must find $k$ points $C$ that
minimize the cost function\vspace{-1ex}

\[
  f(C) = \sum_{p\in P} \textrm{min}\{ ||p-c||, c \in C \}
\]

Solving this with Newton's Method requires computing the Jacobian and
Hessian of $f$.  The cost function is easily written with nested
\texttt{map} and \texttt{reduce} operations, for which the Hessian can
be computed by nesting \texttt{vjp} inside of \texttt{jvp}.  Note that
while
\[
  f : \mathbb{R}^{k\times d} \rightarrow \mathbb{R}
\]
and so needs reverse-mode \texttt{vjp}, the differentiated function
\[
  \textrm{vjp}~f : \mathbb{R}^{k\times d} \rightarrow \mathbb{R} \rightarrow \mathbb{R}^{k\times d}
\]
is suitable for use with forward-mode \texttt{jvp} (the $\mathbb{R}$
argument is set to $1$).  Further, as the Hessian for $f$ has nonzero
entries only along the diagonal, we need only a single invocation of
\texttt{jvp} to compute it, returning exactly this diagonal.  This
shows how \texttt{jvp}/\texttt{vjp} allows the user to exploit
sparsity.  Further, the nesting of \texttt{jvp}/\texttt{vjp} allows us
to avoid duplicating the expensive search operation in the cost
function, which would not be the case if we were limited to dedicated
constructs for Jacobian and Hessian computation.

$k$-means clustering can also be solved using a method based on
repeatedly calculating histograms~\cite{histo-sc20}, which can be
implemented using Futhark's dedicated histogram construct.  The code
that is generated by AD consists of \texttt{map}s that perform
semantically equivalent histogram-like accumulator updates.
Essentially, using AD leads to basically the same code one would write
by hand.  The main difference is that in the hand-written code, some
of the updates have are manually sequentialized, while the AD version
maximizes parallelism to degree that is detrimental for some datasets.

In PyTorch we must take care when computing the all-pairs Euclidean
distance from data points to the centers. Expressing this computation
with broadcasting suffers from massive memory overhead. We get around
this by explicitly distributing the quadratic terms in the Euclidean
distance.  The AutoGrad module computes the Jacobian and Hessian,
instead of in one go as we can when nesting \texttt{vjp}/\texttt{jvp}.

We benchmark with two qualitatively different datasets, with results
in \cref{tab:kmeans}.  When the histograms benefit from the
optimisations discussed in~\cite{histo-sc20}, the hand-written
implementation has a speedup of $4\times$ over our AD approach,
while there is no significant difference if they cannot. Our AD
approach is slightly faster than PyTorch (as high as $1.22\times$
on \textbf{A100} and $1.18\times$ on \textbf{2080Ti}).


\subsection{Case Study 2: Sparse $K$-means clustering}
\label{sec:sparse-kmeans}

\begin{table}
  \centering
  \begin{tabular}{lc|rr|r}
                                                                       & & \multicolumn{2}{c|}{\textbf{Futhark}} & \\\
                                                                       & Workload & \textbf{Manual} & \multicolumn{1}{c|}{\textbf{AD}} & \textbf{PyTorch} \\\hline
 \multirow{2}{*}{\rotatebox[origin=c]{90}{\scriptsize\textbf{A100}}}   & movielens & $61ms$ & $152ms$ & $61223ms$ \\
                                                                       & nytimes   & $83ms$ & $300ms$ & $226896ms$ \\
                                                                       & scrna     & $156ms$ & $579ms$ & $367799ms$\\
  \end{tabular}
  \caption{Performance measurements on \textbf{A100} for sparse $k$-means
    for three NLP workloads, implemented both manually in Futhark,
    with AD in Futhark, and with AD in PyTorch.}
  \label{tab:sparse-kmeans}
\end{table}


We have implemented and tested a sparse formulation of k-means, that uses a 
dense representation for centroids and a sparse representation of data.
Futhark implementations are relatively straightforward adaptations of the
dense $K$-means to use the CSR format instead.

The PyTorch implementation uses the COO format because AD (jvp and hvp)
currently raises runtime errors with CSR format.\footnote{We have an
open (unresolved) query on the PyTorch discord asking if CSR will be
supported in the future.}
To efficiently compute the cost function we expand the all pairs norm
between data $d$ and centroids $c$: $||d_i - c_j||^2 = d_i^2 + c_j^2 - 2 * d_ic_j^T$.
In expanded form, all terms can be computed using vectorized operations. 
Due to the array (tensor) presentation $d_i^2$ and $d_i c_j^T$
are computed using PyTorch specialized sparse module, in particular,
the (sparse) matrix product uses {\tt sparse.mm} which allows sparse
gradient computation (but also forces the COO representation).

\cref{tab:sparse-kmeans} shows the runtime results for \textbf{A100}
on three publicly available NLP workloads for $k=10$.
Our AD is slower than the manual code by a factor between
$2.5-3.7\times$ because the generalized-histogram
implementation of the manual code uses a multi-pass technique
that allows the updates to fit in the L2 cache~\cite{histo-sc20}. 
PyTorch AD is more than $400\times$ slower than our AD, which
we do not know how to explain.

\subsection{Case Study 3: GMM}
\label{sec:gmm}

\begin{table}
\begin{subtable}{0.5\textwidth}
\centering
\begin{tabular}{lccc|lccc}
               & $n$ &  $d$& $K$  &               & $n$   & $d$ & $K$ \\\hline
$\mathbf{D}_0$ & 1k  & 64  & 200  & $\mathbf{D}_3$ & 10k & 64  & 25  \\
$\mathbf{D}_1$ & 1k  & 128 & 200  & $\mathbf{D}_4$ & 10k  & 128 & 25 \\
$\mathbf{D}_2$ & 10k & 32  & 200  & $\mathbf{D}_5$ & 10k  & 128 & 200
\end{tabular}
\caption{Parameters of the ADBench datasets used for benchmarking. The
  corresponding datasets may be found on the ADBench GitHub
  repository (\url{https://github.com/microsoft/ADBench}).}
\label{tab:gmm-data}
\end{subtable}
\begin{subtable}{0.5\textwidth} 
\centering
\begin{tabular}{l|cccccc}
                       & $\mathbf{D}_0$ & $\mathbf{D}_1$ & $\mathbf{D}_2$ & $\mathbf{D}_3$  & $\mathbf{D}_4$ & $\mathbf{D}_5$\\ \hline
\textbf{PyT. Jacob. (ms)}          & $7.6$   & $19.2$  & $19.2$   & $7.7$   & $18.3$  & $67.7$ \\
\textbf{Fut. Speedup ($\times$)}   & $1.85 $ & $2.18$  & $1.45 $  & $1.81$  & $1.89$  & $0.87$ \\
\textbf{PyT. Overhead ($\times$)}  & $2.64 $ & $5.28$  & $2.45 $  & $3.09$  & $4.04$  & $2.46$ \\
\textbf{Fut. Overhead ($\times$)}  & $2.34$  & $2.20$  & $2.24 $  & $2.0 $  & $2.98$  & $3.18$  
\end{tabular}
\caption{GMM benchmark results on the \textbf{A100} system with 64-bit
  floats.  {\bf Fut.} and {\bf PyT.} refer to Futhark and PyTorch.
  {\bf PyT. Jacob.} is the time to compute the full
  Jacobian of the objective function in PyTorch. }
\label{tab:gmm-a100}
\end{subtable}

\begin{subtable}{0.5\textwidth} 
\centering
\begin{tabular}{l|cccccc}
                       & $\mathbf{D}_0$ & $\mathbf{D}_1$ & $\mathbf{D}_2$ & $\mathbf{D}_3$  & $\mathbf{D}_4$ & $\mathbf{D}_5$\\ \hline
\textbf{PyT. Jacob. (ms)}         & $20.2$ & $57.9$ & $98.8$  & $31.1$ & $72.5$ & $\color{red}\varnothing$ \\
\textbf{Fut. Speedup ($\times$)}   & $4.5$  & $5.29$ & $4.78 $ & $7.31$ & $6.20$ & $\color{red}\varnothing$ \\
\textbf{PyT. Overhead ($\times$)} & $2.28$ & $2.41$ & $3.07$  & $3.20$ & $2.62$ & $\color{red}\varnothing$ \\
\textbf{Fut. Overhead ($\times$)}  & $2.8$  & $2.5$  & $3.2$   & $2.6$  & $3.5$  & $3.27 $       
\end{tabular}
\caption{GMM benchmark results on the \textbf{2080Ti} system with
  32-bit floats. The PyTorch implementation ran out of memory on the
  $\mathbf{D}_5$ dataset, indicated with $\color{red}\varnothing$. The
  Futhark Jacobian-runtime was $96.7ms$ on $\mathbf{D}_5$.}
\label{tab:gmm-2080ti}
\end{subtable}
\vspace{-3pt}
\caption{Benchmarking results for the GMM case study.}
\end{table}

To evaluate the parallelism-preservation of our AD transformation, we
compile the GMM benchmark from the ADBench suite to parallel CUDA. We
compare against ADBench's implementation of GMM in PyTorch (also run
on CUDA), which we have improved (by a $>10\times$ factor) by
vectorizing all comprehensions. We
benchmark on a selection of 1,000 and 10,000-point datasets from ADBench
featuring a variety of $d$ (the dimensionality of the input data) and
$K$ (the number of Gaussian distributions used in the model) values,
see \cref{tab:gmm-data}. We observed that the
runtime of the primal program is dominated by matrix multiplication
($\sim70\%$).

Matrix multiplication is a primitive in PyTorch
\cite{paszke2019pytorch}; we reasonably expect that
differentiation of matrix multiplication may be readily implemented
very efficiently. In Futhark there are no such primitives: 
matrix multiplication is written with maps, whose differentiation
yield accumulators, which are further optimized as described
in \cref{subsec:opt-mat-mul}.

The benchmark results are shown in Tables \ref{tab:gmm-a100} and
\ref{tab:gmm-2080ti} for the \textbf{A100} and \textbf{2080Ti}
systems, respectively. On the \textbf{A100} 64-bit floats were used;
on the \textbf{2080Ti} system 32-bit floats were used due to the
limited double-precision performance of the system. 
The results on \textbf{2080Ti} demonstrate a significant speedup
of Futhark over PyTorch (as high as $7.3\times$) in the Jacobian 
runtimes on all inputs. Futhark also gets the upper hand on  
\textbf{A100}, albeit the average speedup is smaller ($1.68\times$).

\subsection{Case Study 3: LSTM}
\label{sec:lstm}

\begin{table}
\centering
\addtolength{\tabcolsep}{-3pt}
\begin{tabular}{llr|rr|rrc}
                                                         &                                     &                          & \multicolumn{2}{c}{\textbf{Speedups} ($\times$)}        & \multicolumn{3}{|c}{\textbf{Overheads} ($\times$)}  \\
                                                         &                                     & \textbf{PyT. Jacob.} & \textbf{Fut.}   &\multicolumn{1}{c|}{\textbf{cuDNN}}   & \textbf{PyT.}   & \textbf{Fut.}     & \textbf{cuDNN}    \\ \hline
\multirow{2}{*}{\rotatebox[origin=c]{90}{\scriptsize\textbf{A100}}}   & \multicolumn{1}{|c}{$\mathbf{D_0}$} & \multicolumn{1}{c|}{$51.9ms$}       &  3.1 &\multicolumn{1}{c|}{14.0}     &  2.6            & 2.0               & 2.8                    \\
                                                                          & \multicolumn{1}{|c}{$\mathbf{D_1}$} & \multicolumn{1}{c|}{$713.7ms$}  &  3.0 & \multicolumn{1}{c|}{25.5}    &  3.6            & 4.0               &  1.7                 \\\hline
\multirow{2}{*}{\rotatebox[origin=c]{90}{\scriptsize\textbf{2080Ti}}} & \multicolumn{1}{|c}{$\mathbf{D_0}$} & \multicolumn{1}{c|}{$76.8ms$}       &  3.1 &\multicolumn{1}{c|}{8.3}      &  2.6            & 2.6               &  3.2                   \\
                                                                          & \multicolumn{1}{|c}{$\mathbf{D_1}$} & \multicolumn{1}{c|}{$1175.0ms$} &  3.2 & \multicolumn{1}{c|}{12.8}    &  2.5            & 4.0               &  2.5                 \\\hline
\end{tabular}
\caption{LSTM measurements on datasets
  $\mathbf{D_0}: (bs,n,d,h) = (1024, 20, 300, 192)$ and $\mathbf{D_1}:
  (bs,n,d,h) = (1024, 300, 80, 256)$. \textbf{Py.T}, \textbf{Fut.}, and
  \textbf{cuDNN} refer to the PyTorch, Futhark, and the cuDNN-based
  \lstinline{torch.nn.LSTM} implementations.}
\label{tab:lstm}
\addtolength{\tabcolsep}{3pt}
\end{table}\vspace{-1ex}

Long Short-Term Memory (LSTM) \citep{sak2014long} is a type of
recurrent neural network architecture popular in named entity
recognition and part-of-speech tagging
\citep{chiu2016named}\citep{ronning2018sluice}.  We benchmark two LSTM
networks with hyperparameters common in natural language processing.
The LSTM's input dimensions ($d$) are 80 and 300, respectively. These
correspond to GloVe embeddings \citep{pennington2014glove}. The
recurrent units have hidden dimensions ($h$) 192 and 256
\citep{chiu2016named}. The sequence lengths ($n$) are 20 and 300.  The
lengths are close to the average sentence length for the Penn Treebank
dataset \citep{marcus1993building} and IMDB dataset
\citep{maas2011learning}. Batch sizes $(bs)$ were chosen to
fully saturate the GPUs.

Both the Futhark and PyTorch implementations are based on the
architecture in \citep{sak2014long}. We also compare against PyTorch's
\lstinline{torch.nn.LSTM} class, which wraps the NVIDIA cuDNN LSTM
implementation \citep{chetlur2014cudnn}; note that this implementation
is hand-written/optimized and features manual differentiation. All
benchmarks were on 32-bit floating points and were initialized with
the same parameter data, which was generated by
\lstinline{torch.nn.LSTM}. The results are shown in \cref{tab:lstm}.

Futhark is about $3\times$ faster than PyTorch on both \textbf{2080Ti}
and \textbf{A100} systems. As with GMM, LSTM is dominated by
matrix-multiplication computations.  Neither AD-based implementation
is competitive against the manual cuDNN-based implementation---we
suspect that \lstinline{torch.nn.LSTM} switches the computation to
tensor cores ($F16$) internally, because the speedups over Futhark
are well correlated with the ratio between peak $F32$ and $F16$
performance of the two systems.
%

\section{Related Work}
\label{sec:relatedwork}

Reverse-mode differentiation of \texttt{reduce} and \texttt{scan} is
discussed in \cite{10.1145/3471873.3472975}.  Our rule for
\texttt{reduce} is similar, but we handle \texttt{scan} differently.
Our approach is less efficient for complex operands because we
manifest the Jacobian matrices, but more efficient for single-value
operands on GPUs as it requires less shared memory to implement the
derived \texttt{scan} operator.  Neither our \texttt{scan} rule nor
the one from \cite{10.1145/3471873.3472975} is asymptotics-preserving
in the general case, but they are for most \texttt{scan}s that occur
in practice.

$\widetilde{F}$ is a functional array language that supports nested
parallelism.  Its AD implementation uses the forward mode, along with
a handful of rewrite rules that allow it to exploit sparsity in some
cases~\cite{10.1145/3341701}.

Dex is a recent language built specifically to support efficient AD.
Empirical benchmarks for AD in Dex have not yet been published, but we
can compare their approach~\cite{10.1145/3473593}.  In contrast to our
conventional ``monolithic'' approach where reverse-mode AD is a
transformation completely distinct from forward-mode, Dex uses a
technique where the program is first ``linearized'', producing a
linear map, after which this linear map is then ``transposed'',
producing the adjoint code.  Like Dex, we do not support recursion or
AD of higher-order functions.  Dex does not make direct use of a
``tape'' in the classical sense, but instead constructs arrays of
closures followed by defunctionalization.  The actual run-time data
structures will conceptually consist of \emph{multiple} tapes in the
form of multidimensional irregular arrays.  Dex does not yet make use
of checkpointing, or optimization of particular accumulation patterns
as in \cref{subsec:opt-mat-mul}.

Enzyme shows the advantage of performing AD after standard compiler
optimisations has simplified the program~\cite{enzymeNeurips}.  Like
Enzyme, we also apply our AD transformation on a program that has
already been heavily optimized by the compiler.  But where Enzyme is
motivated by performing AD on a post-optimization low-level
representation, our work takes advantage of \emph{both} pre-AD
optimization, as well as the information provided by high-level
parallel constructs.
Enzyme has also been applied to GPU
kernels~\cite{10.1145/3458817.3476165}.  We achieve equivalent
performance, but our approach is not based on differentiating single
kernels---indeed, the GPU code we generate for a differentiated
program may have a significantly different structure than the original
program.  For example, the optimized adjoint code for a matrix
multiplication requires \emph{two} matrix multiplications, each its
own kernel, as in the LSTM and GMM benchmarks.

Recent AD work on OpenMP details an approach to reverse-mode AD for
non-nested parallel loops \cite{huckelheim2021source}. As with
\kw{map}, the chief difficulty is handling adjoint updates to free
variables in the loop. Two approaches are used: a) the variable is
copied across threads and per-thread updates to the adjoint will be
combined at loop termination or b) the variable is shared across
threads and updates are handled via atomic operations. Copying free
variables across threads does not in general preserve the asymptotics
of the original program when the free variables are arrays; shared
variables avoid this issue, but instead can result in significant
sequentialization. Due to our redundant recomputation technique
(avoiding the difficulty of passing a tape across scopes), our
implementation supports and preserves nested parallelism in the
original program. Hence, while our implementation is also realized by
atomic operations, we are able to exploit nested parallelism to
identify and reduce some updates into a single atomic update and
thereby leverage benefits from both approaches.

ML practitioners use tools such as PyTorch~\cite{paszke2019pytorch}
that incorporate AD.  These are less expressive than our language and
do not support true nested parallelism, but instead require the
program to use flat (although vectorized) constructs.  On the other
hand, they can provide hand-tuned adjoints for the primitives they do
support. JAX is another such example; it supports automatic
differentiation of pure Python code and just-in-time (JIT) compilation to
XLA HLO~\cite{frostig2018compiling,jax2018github}. However, JAX
does not support reverse-mode differentiation of their loop
primitive (\texttt{fori\_loop}) when the iteration count of the loop
isn't statically known. JAX's JIT approach requires re-interpretation
(by Python) and compilation (by XLA) when functions are called on
arguments with so far unseen shape/underlying type, but this approach
may enable better code specialization than our entirely ahead-of-time
approach.

Reverse AD has also been implemented in DSLs aimed at stencil
computations~\cite{stencil-ad}. The challenge here is to combine AD
with loop transformations such that the resulted code is a stencil
itself, and thus it can be optimized with the repertoire of existent
optimizations.  AD has also been described for tensor languages that
support only constrained forms of loops, which in particular has the
benefit of not requiring the use of
tapes~\cite{bernstein2020differentiating}.

\section{Conclusions}
We have presented a fully operational compiler implementation of
both reverse and forward mode AD in a nested-parallel functional
language. Our experimental evaluation shows that our mapping
of parallel construct and our technique of implementing the tape
based on redundant computation is practically effective, and
competitive with both (1) well-established frameworks that
encompass more specialized languages such as PyTorch, and
with (2) newer research efforts aimed at a lower-level language,
such as Enzyme.


\bibliography{ad-sc22}

\end{document}